\documentclass[11pt]{article}
\usepackage{epsfig} 
\usepackage{graphics}
\newcommand{\sect}[1]{ \section{#1} \setcounter{equation}{0} } 

\newcommand{\half}{\mbox{\small{$\frac{1}{2}$}}}

\newcommand{\twopithree}{\mbox{\small{$\frac{2\pi}{3}$}}} 
\newcommand{\MSbar}{\overline{\mbox{MS}}} 
\newcommand{\MSbars}{\overline{\mbox{\footnotesize{MS}}}}
\newcommand{\Nc}{N_{\!c}}
\newcommand{\Nf}{N_{\!f}}
\newcommand{\NA}{N_{\!A}}
\newcommand{\bare}{\mbox{\footnotesize{o}}}
\newcommand{\Dslash}{D \! \! \! \! /}

\begin{document}

\title{Three loop effective potential for $\langle \frac{1}{2} { A_\mu^a }^2 
\rangle$ in the Landau gauge in QCD}

\author{J.A. Gracey, \\ Theoretical Physics Division, \\ 
Department of Mathematical Sciences, \\ University of Liverpool, \\ P.O. Box 
147, \\ Liverpool, \\ L69 3BX, \\ United Kingdom.} 
\date{}

\maketitle 

\vspace{5cm} 
\noindent 
{\bf Abstract.} We apply the Local Composite Operator method to construct the
three loop effective potential for the dimension two operator 
$\frac{1}{2} { A_\mu^a }^2$ in the Landau gauge in Quantum Chromodynamics. For
$SU(3)$ we show that the three loop value of the effective mass of the gluon is
similar to the two loop estimates when the number of massless quarks is  
strictly less than five for $SU(3)$.

\vspace{-16.0cm}
\hspace{13.2cm}
{\bf LTH 1309}

\newpage 

\sect{Introduction.}

In the last quarter century it has become accepted that the behaviour of the
Landau gauge gluon propagator in Quantum Chromodynamics (QCD), as computed on
the lattice, is not of the fundamental form, \cite{1,2,3,4,5,6,7,8,9}. In other
words it differs from that of the photon in Quantum Electrodynamics which has a 
massless pole with respect to the momentum. Instead the gluon propagator is 
bounded with no singularity at any momentum and moreover freezes to a non-zero 
value. In the high energy limit, if $p$ is the momentum, then the propagator 
has a $1/p^2$ asymptotic behaviour which is consistent with the Lagrangian 
construction that gluons behave as effectively free particles analogous to a 
non-abelian massless photon. That the gluon propagator freezes at zero momentum
indicates that some non-zero mass scale is associated with the infrared 
properties of the field. This is loosely termed a gluon mass but not because 
one can identify an isolated state with a massive {\em fundamental} field. 
Clearly the presence of such a mass scale, as opposed to a canonical mass, 
requires further understanding from a theoretical point of view. There has been
substantial progress to this end since the early lattice observations of 
\cite{1} and subsequent confirmations. See, for example, 
\cite{2,3,4,5,6,7,8,9}. One prior idea centred on Gribov's observation, 
\cite{10}, that to globally uniquely fix a covariant gauge in a non-abelian 
gauge theory such as QCD requires a restriction on the path integral, 
\cite{10}. In the Landau gauge this introduces a mass scale, the Gribov mass, 
that satisfies a gap equation equating it to a non-perturbative function of the
coupling constant. The fundamental structure of the gluon propagator is 
modified as well. At large momenta it behaves like the propagator of a massless 
field with a $1/p^2$ asymptote. By contrast at low momenta the non-zero Gribov 
mass modifies the propagator in that it is bounded with no singularity and 
vanishes at zero momenta. Evidently this is out of line with the low momenta 
lattice results. Several modifications to Gribov's construction have been 
introduced such as Zwanziger's programme to localize the Gribov operator to 
produce a local renormalizable Lagrangian, 
\cite{11,12,13,14,15,16,17,18,19,20,21}, that involves additional spin-$1$ 
ghost fields. While this maintained the suppressed gluon propagator in the 
infrared it was the foundation for a subsequent extension. A dimension two 
gluon mass operator was also included and studied at length in \cite{22,23,24}.
The localized Gribov-Zwanziger Lagrangian with the extra operator is also
renormalizable. In this particular modification of the Gribov construction 
various composite operators of the fields can develop non-zero vacuum 
expectation values. Consequently for certain configurations the resulting gluon
propagator was much more in line with lattice data in that they froze to a 
non-zero finite value.

The idea of treating the QCD Lagrangian by appending a simple gluon mass 
operator without spontaneous symmetry breaking predates Gribov's seminal work.
In the mid-1970s Curci and Ferrari, \cite{25}, investigated a nonlinear gauge
fixing of Yang-Mills theory where a BRST invariant dimension two gluon and
ghost mass operator was included. This idea fell out of fashion for a while due
to the loss of nilpotency of the BRST charge, 
\cite{26,27,28,29,30,31,32,33,34}. More recently there has been a revival of 
interest in the model and a gluon mass term. For instance, it was shown in 
\cite{35,36,37,38,39} that one could connect the mass parameter of the gluon 
with Gribov copies that arise in the global gauge fixing, \cite{10}. In other 
words one can interpret the mass as an extra gauge parameter reflecting the 
effect of the copies so that the infrared properties of Yang-Mills theory could
be described by a Lagrangian with a gluon mass term. This perspective was 
tested out at one loop in \cite{37,40,41} where the gluon mass model was used 
to calculate the gluon propagator analytically and then fitted to lattice data.
There was a reasonable quantitative agreement over virtually all momenta. The 
inclusion of the two loop corrections was carried out in \cite{42} with a 
perceptible improvement on the one loop fit over all momenta. This lends
support to the idea that a massive gluon could be a useful tool to model 
infrared phenomena in QCD. One observation was that the mass was of the order 
of $350$ MeV. This was not inconsistent with recent estimates of the mass gap 
such as fitting models of the gluon propagator to lattice data, \cite{43}, or 
via functional renormalization group ideas, \cite{44}. Indeed the former 
article also suggested that a running gluon mass might be sufficient to 
circumvent the problem of Gribov copies. That such a consistent mass scale 
emerges in Yang-Mills theory or QCD from different techniques provides evidence
that this is a particular infrared property.

While a two loop analysis was carried out in \cite{42} if it is accepted that 
there is a gluon mass scale then the various Lagrangian field theory approaches
ought to be refined by extending them to three loops. However computing the 
three loop corrections to the gluon propagator in a model with a massive gluon 
over all momenta along the lines of \cite{42} is not currently possible on 
technical grounds. For instance, the three loop massive $2$-point master 
Feynman integrals are not available analytically for all momenta. Their values 
are necessary for the final stage in applying the Laporta algorithm, \cite{45}, 
which is the main programme for evaluating the various $2$-point functions.
Instead, as an alternative approach to gain three loop insight, it is possible 
to study the effect of three loop corrections on the gluon mass gap by another 
technique. In \cite{46} a method was developed and subsequently refined, 
\cite{47}, to compute the effective action and associated effective potential 
of dimension two operators in a quantum field theory. Termed the Local 
Composite Operator (LCO) method it was used at one and two loops in
\cite{46,47} to study the $SU(N)$ Gross-Neveu model, \cite{48}. Although this 
is a two dimensional renormalizable theory it shares various properties with 
QCD that include asymptotic freedom, dynamical mass generation and the 
existence of a mass gap, \cite{48}. In the Gross-Neveu model the latter has a 
significant property in that the $N$ dependence of the mass gap is known 
{\em exactly}, \cite{49,50}, providing a way to benchmark the LCO method. 
Indeed the mass estimates determined in \cite{46,47} were in close agreement 
with the exact mass gap expression for a large range of $N$. Buoyed by this 
observation the LCO method was applied to Yang-Mills theory at two loops in 
\cite{51} and later to QCD, \cite{52}, at the same order. The gluon mass 
estimates given in \cite{51,52} were roughly $2\Lambda_{\MSbars}$ where the 
definition of the effective mass was the vacuum expectation value of the scalar
colour singlet field that arises as part of the LCO formalism. As the mass gap 
value of \cite{51,52} in the early 2000s predated more recent estimates it was 
difficult then to assess its consistency with other approaches.

Given the recent interest in gluon mass determinations from the lattice,
functional renormalization group method and other approaches such as the
Schwinger-Dyson technique, it is the purpose of this article to extend the LCO 
analysis of Yang-Mills theory and QCD to three loops. To be able to go to this 
loop order has become possible now with advances in techniques to evaluate high 
order Feynman integrals. For instance, we will make use of the {\sc Forcer} 
algorithm, \cite{53,54}, as well as the Laporta algorithm, \cite{45}, to 
respectively evaluate the four and three loop Feynman integrals that are 
necessary to construct the three loop effective potential of the dimension two 
gluon mass operator or equivalently the colour singlet scalar that condenses to
produce the gluon effective mass. The advantage of the LCO method is that it 
provides a quantum field theoretic foundation. By this we mean that the 
effective potential satisfies a homogeneous renormalization group equation and 
the effective action is linear in the source field associated with the colour 
singlet scalar, \cite{46,47,51}. This leads to an effective potential with a 
well-established energy interpretation, \cite{46,47,51}, that means a 
well-defined mass can be extracted from the absolute minimum of the potential 
which is not the perturbative one. Moreover the underlying Lagrangian, which is
modified by the presence of the colour singlet scalar, is renormalizable. Once 
the effective potential has been computed at three loops for an arbitrary 
colour group, we will extract estimates for the effective gluon mass. The 
definition of the mass we propose to use here will differ from that of the 
earlier studies of \cite{51,52} which simply involved the vacuum expectation 
value of the scalar field. Instead we will define the mass as the coefficient 
of the gluon mass operator in the LCO derived Lagrangian in the 
non-perturbative vacuum. It will turn out that mass gap estimates derived with 
this definition will be more stable with respect to loop corrections as well as
more in keeping with values deduced by other methods.

The article is organized as follows. We recall the formal aspects of the LCO
method in Section $2$ in the Yang-Mills theory context including the derivation
of the modified Lagrangian. Section $3$ is devoted to the computation of the 
three loop effective potential that includes the calculation of the underlying
fundamental LCO parameter from the evaluation of a four loop massless $2$-point
function. Aspects of the renormalization of the LCO Lagrangian are also 
discussed in Section $3$. The three loop effective potential is analysed in 
Section $4$ for both $SU(2)$ and $SU(3)$ for a range of quark flavours which 
lead to estimates of the effective gluon mass. Concluding remarks are provided 
in Section $5$ with an Appendix recording expressions for all the relevant 
perturbative quantities needed to determine the potential, as well as the 
potential itself, for a general colour group.

\sect{Formalism.}

We devote this section to reviewing the basics of the LCO method as applied to 
QCD, \cite{51,52}, in the Landau gauge. By way of introducing conventions we 
recall the QCD Lagrangian is
\begin{equation} 
L_{\bare} ~=~ -~ \frac{1}{4} G_{{\bare}\,\mu\nu}^a G_{\bare}^{a \, \mu\nu} ~-~ 
\frac{1}{2\alpha_{\bare}} (\partial^\mu A^a_{{\bare}\,\mu})^2 ~-~ 
\bar{c}_{\bare}^a \partial^\mu D_{{\bare}\,\mu} c_{\bare}^a ~+~ 
i \bar{\psi}_{\bare}^{iI} {\Dslash}_{\bare} \psi_{\bare}^{iI} 
\label{qcdlag}
\end{equation} 
where the index ranges are $1$~$\leq$~$i$~$\leq$~$\Nf$,
$1$~$\leq$~$a$~$\leq$~$\NA$ and $1$~$\leq$~$I$~$\leq$~$\Nc$, $\Nf$ is the 
number of massless quarks, $\NA$ is the dimension of the adjoint representation
and $\Nc$ is the number of colours or equally the dimension of the fundamental 
representation. All entities in (\ref{qcdlag}) are bare as denoted by the 
subscript ${}_{\bare}$. The field strength and covariant derivatives are
\begin{eqnarray} 
G_{\mu\nu}^a &=& \partial_\mu A^a_\nu ~-~ \partial_\nu A^a_\mu ~-~
g f^{abc} A^b_\mu A^c_\nu \nonumber \\
D_\mu c^a &=& \partial_\mu c^a ~-~ g f^{abc} A^b_\mu c^c ~~,~~ 
D_\mu \psi^{iI} ~=~ \partial_\mu \psi^{iI} ~+~ i g T_{IJ}^a A^a_\mu \psi^{iJ}
\end{eqnarray} 
where $g$ is the gauge coupling constant. Although we have included the gauge
parameter $\alpha$ in (\ref{qcdlag}) it will henceforth be set to the Landau 
gauge value of zero in all subsequent calculations. We focus on this particular
gauge since it is the one that has been most widely studied over many years,
using lattice field theory and Schwinger-Dyson methods, meaning that the gluon 
mass has been estimated in it more than any other gauge. Having recalled the 
core QCD Lagrangian we now focus on the incorporation of the dimension $2$ 
operator ${\cal O}$~$=$~$\frac{1}{2} { A_\mu^a }^2$ into the LCO formalism. 
This is achieved by considering the path integral where the operator is 
included with a source $J$ leading to the additional term $J {\cal O}$ in the 
Lagrangian which will then have an associated path integral generating 
functional $W[J]$. Therefore the starting point to treat ${\cal O}$ in the LCO 
method is given by the functional, \cite{51}, 
\begin{equation}
e^{-W[J_{\bare}]} ~=~ \int {\cal D} A^\mu_{\bare} {\cal D} \psi_{\bare} 
{\cal D} \bar{\psi}_{\bare} {\cal D} c_{\bare} {\cal D} \bar{c}_{\bare} \,
\exp \left[ \int d^d x \left( L_{\bare} ~-~ \frac{1}{2} 
J_{\bare} A_{{\bare} \, \mu}^{a \, 2} ~+~ 
\frac{1}{2} \zeta_{\bare} J_{\bare}^2 \right) \right]
\end{equation}  
where everything is expressed in terms of bare quantities. In addition to the
linear source term there is a quadratic one. This is necessary to ensure
renormalizability which can be seen in several ways. The simplest is through
dimensional analysis, \cite{51,52}. As the gluon field has dimension $1$ then 
$J$ has dimension $2$ and therefore a quadratic term in $J$ is necessary in 
four dimensions. If such a term was not present then the correlation function 
of $J$, which is clearly divergent, would not have an available counterterm to
allow the consistent redefinition of the bare Lagrangian in terms of 
renormalized variables. Separately the additional parameter $\zeta$ has been 
introduced in order to ensure that the renormalization group equation for 
$W[J]$ is homogeneous, \cite{46,47,51}. The actual form of the renormalized 
parameter will be calculated later via the renormalization group construction 
and will be a perturbative function of the coupling constant, \cite{46,47,51}.

As the bare Lagrangian $L_{\bare}$ has already been given the renormalized 
quantities are defined in the canonical way by
\begin{eqnarray}
A^a_{{\bare}\,\mu} &=& \sqrt{Z_A} A^a_\mu ~~,~~
\bar{c}_{\bare}^a ~=~ \sqrt{Z_c} c^a ~~,~~
\psi_{\bare}^{iI} ~=~ \sqrt{Z_\psi} \psi^{iI} \nonumber \\
g_{\bare} &=& \mu^\epsilon Z_g g ~~,~~
\alpha_{\bare} ~=~ \frac{Z_A}{Z_\alpha} \alpha 
\label{rencon}
\end{eqnarray} 
where $\mu$ is the scale introduced to ensure the coupling constant remains
dimensionless in dimensional regularization in $d$~$=$~$4$~$-$~$2\epsilon$,
which we use throughout, and the source renormalization is achieved via, 
\cite{46,47,51},
\begin{equation} 
J_{\bare} ~=~ \frac{Z_m}{Z_A} J ~~~,~~~ 
\zeta_{\bare} J_{\bare}^2 ~=~ \left( \zeta ~+~ \delta \zeta \right) J^2 ~.
\label{renconj}
\end{equation} 
The quantity $\delta \zeta$ should be regarded as a counterterm and $Z_m$ is
the renormalization constant associated with the renormalization of the 
dimension two operator ${\cal O}$ or equivalently the gluon mass. Consequently
we have 
\begin{equation}
e^{-W[J]} ~=~ \int {\cal D} A_\mu {\cal D} \psi {\cal D} \bar{\psi} {\cal D} c 
{\cal D} \bar{c} \, \exp \left[ \int d^d x \left( L ~-~ \frac{1}{2} Z_m J 
A_\mu^{a \, 2} ~+~ \frac{1}{2} ( \zeta + \delta \zeta ) J^2 \right) \right] ~. 
\end{equation}  
for the generating functional for $J$ in terms of renormalized quantities.

Next we recall the procedure to find $\zeta$ as an explicit function,
\cite{46,47,51}. This will be achieved from the renormalization group 
properties of $W[J]$ which satisfies
\begin{equation}
\left[ \mu \frac{\partial ~}{\partial \mu} ~+~ \beta(a) \frac{\partial ~}
{\partial a} ~-~ \gamma_m(a) \int_x J \frac{\delta ~}{\delta J} ~+~
\mu \frac{\partial \zeta}{\partial \mu} \frac{\partial ~}{\partial 
\zeta} \right] W[J] ~=~ 0 
\end{equation} 
where we have set $a$~$=$~$g^2/(16\pi^2)$. From (\ref{renconj}) we can deduce 
that 
\begin{equation}
\mu \frac{\partial \zeta}{\partial \mu} ~=~ 2 \gamma_m(a) \zeta ~+~ 
\delta(a) 
\end{equation}
where $\delta(a)$ is in effect the anomalous dimension of $J$ and is 
determined from the counterterm via, \cite{46,47,51},
\begin{equation}
\delta(a) ~=~ \left[ 2 \epsilon ~+~ 2 \gamma_m(a) ~-~ 
\beta(a) \frac{\partial~}{\partial a} \right] \delta \zeta 
\label{deltadef}
\end{equation} 
in order for $W[J]$ to satisfy a homogeneous renormalization group equation,
\cite{46,47,51}. As it stands presently this renormalization group function 
follows from the standard procedure for treating bare parameters in a 
renormalizable theory. In the context of the LCO method the origin of the 
$\delta \zeta$ counterterm is due to the requirement of the quadratic term in 
$J$ for renormalizability since the $2$-point correlation function of $J$ is 
divergent. This implies that $W[J]$ is not linear in the source which would 
rule out an energy interpretation for the action after a Legendre 
transformation. Equally the parameter $\zeta$ is undetermined at this point. To
resolve this, \cite{46,47,51}, $\zeta$ is chosen to be the solution of the 
first order differential equation
\begin{equation}
\beta(a) \frac{\partial \zeta(a)}{\partial a} ~=~ 2 \gamma_m(a) \zeta(a) ~+~ 
\delta(a) ~. 
\label{zetasoln}
\end{equation}
In other words the running of $\zeta(a)$ is determined from that of the
coupling constant. If one examines the $a$ dependence of the various known
expressions in (\ref{zetasoln}) we note that $\beta(a)$~$=$~$O(a^2)$,
$\gamma_m(a)$~$=$~$O(a)$ and $\delta(a)$~$=$~$O(1)$ implying the leading term 
of $\zeta(a)$ has to be $O(1/a)$. Therefore one formally solves
(\ref{zetasoln}) perturbatively using the ansatz 
\begin{equation}
\zeta(a) ~=~ \sum_{n \, = \, - \, 1}^\infty c_n a^n
\end{equation} 
which produces a {\em unique} function of the coupling constant leading to the
homogeneous renormalization group equation for $W[J]$, \cite{46,47,51},
\begin{equation}
\left[ \mu \frac{\partial ~}{\partial \mu} ~+~ \beta(a) \frac{\partial ~}
{\partial a} ~-~ \gamma_m(a) \int_x J \frac{\delta ~}{\delta J} 
\right] W[J] ~=~ 0 ~.
\end{equation} 
The main consequence of this is that $\Delta$, given by
\begin{equation}
\Delta ~=~ \frac{\delta W[J]}{\delta J} ~, 
\end{equation}
has a well-defined vacuum expectation value and its effective action follows
from
\begin{equation}
\Gamma[\Delta] ~=~ W[J] ~-~ \int_x \, J \Delta ~.
\label{wjinv}
\end{equation}
Consequently the effective action satisfies
\begin{equation}
\left[ \mu \frac{\partial ~}{\partial \mu} ~+~ \beta(a) \frac{\partial ~}
{\partial a} ~+~ \gamma_m(a) \int_x \Delta \frac{\delta ~}{\delta \Delta} 
\right] \Gamma[\Delta] ~=~ 0 
\end{equation} 
which is the standard formal renormalization group equation.

In order to determine the effective action and thereby the effective potential 
we need to compute $W[J]$ and execute the inversion (\ref{wjinv}). For the case
when $W[J]$ is linear in $J$ this is straightforward but for the present case 
the dependence on $J$ is quadratic. The LCO method bypasses this difficulty for
${\cal O}$ by introducing a Hubbard-Stratonovich transformation that introduces
a purely scalar colour singlet field $\sigma$. Within the path integral this is
effected by including unity in a redundant way, \cite{46,47,51}, via 
\begin{equation}
1 ~=~ \int {\cal D} \sigma \, \exp \left( -~ \left[ b_1 \sigma ~+~ 
b_2 A^{a \, 2}_\mu ~+~ b_3 J \right]^2 \right) ~.
\label{hubstrt}
\end{equation} 
The free parameters $b_i$ are chosen in such a way as to cancel off the 
quadratic term in $J^2$ and the interaction of $J$ with the gluon. In the 
exponent of the integrand in (\ref{hubstrt}) there will be six terms but two of
these will be cancelled by choices of $b_2$ and $b_3$ leaving four terms. One 
of these terms will be linear in $J$ and $\sigma$ allowing $b_1$ to be fixed so
that the new expression for $W[J]$ takes the form 
\begin{equation}
e^{-W[J]} ~=~ \int {\cal D} A_\mu {\cal D} \psi {\cal D} \bar{\psi} {\cal D} c 
{\cal D} \bar{c} {\cal D} \sigma \, \exp \left[ \int d^d x \left( L^\sigma ~-~ 
\frac{\sigma J}{g} \right) \right] ~.
\label{wrgelin}
\end{equation}
Each of the remaining three terms of (\ref{hubstrt}) will involve only fields 
and in particular $A^a_\mu$ and $\sigma$. These are absorbed into a 
redefinition of the original action to produce the new Lagrangian
\begin{eqnarray} 
L^\sigma &=& L_{\bare} ~-~ \frac{\sigma^2}{2g^2 \zeta(a) Z_\zeta} ~+~ 
\frac{Z_m}{2 g \zeta(a) Z_\zeta} \sigma A^a_\mu A^{a \, \mu} ~-~ 
\frac{Z_m^2}{8\zeta(a) Z_\zeta} \left( A^a_\mu A^{a \, \mu} \right)^2
\label{lagsig}
\end{eqnarray} 
where
\begin{equation}
Z_\zeta ~=~ 1 ~+~ \frac{\delta \zeta}{\zeta(a)} ~.
\label{Zzdef}
\end{equation}
The consequence is that with (\ref{lagsig}) one can determine $\Gamma[\Delta]$
using standard techniques of perturbation theory allowing one to deduce the 
effective potential. The only difference for QCD is that its Feynman rules have
to be amended by the additional terms in (\ref{lagsig}) of a modified quartic 
gluon vertex as a well as a momentum independent propagator for the $\sigma$ 
field and its $3$-point interaction with the gluon.

\sect{Determination of the effective potential.}

Having concentrated on the formalism that allows us to extract the effective
potential for the Landau gauge gluon mass operator we devote this section to
the actual computations that lead to the three loop expression. As is clear 
from the previous section various renormalization group functions are required 
to find $\zeta(a)$ such as $\beta(a)$, $\gamma_m(a)$ and $\delta(a)$. Moreover 
as explained these are needed at the four loop level. As the QCD Lagrangian has 
been renormalized to five loops now in the $\MSbar$ scheme, \cite{55,56,57,58},
the only renormalization constants that are needed to extend the two loop 
effective potential of \cite{51,52} are $Z_m$ and the counterterm for $\zeta$. 
The former is in fact already available at five loops in the $\MSbar$ scheme in
the Landau gauge via a simple Slavnov-Taylor identity. It was shown in 
\cite{59} originally and subsequently observed in \cite{60} via a three loop 
calculation that $Z_m$ is not independent but related to $Z_A$ and $Z_c$ in the
Landau gauge. This identity was later verified in more detail in \cite{61}.
Indeed similar Slavnov-Taylor identities have been constructed for the BRST 
invariant gluon mass operators in two nonlinear gauges. These are the 
Curci-Ferrari, \cite{25}, and maximal abelian gauges, \cite{62,63,64}, with the
identities being discussed in \cite{65,66} and \cite{67,68} respectively. So 
$\gamma_m(a)$ in the Landau gauge can be deduced from the combination 
$\gamma_A(a)$~$+$~$\gamma_c(a)$.

{\begin{figure}[ht]
\begin{center}
\includegraphics[width=5.7cm,height=3.5cm]{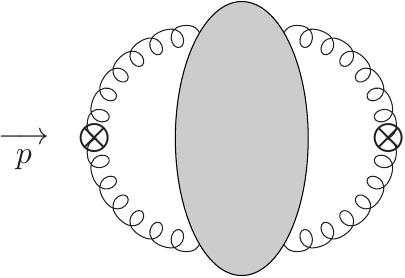}
\end{center}
\caption{Correlation function for the operator source $J$ or ${\cal O}$ where
$\otimes$ denotes the insertion of $J$ or ${\cal O}$.}
\label{figcorj}
\end{figure}}

Therefore it only remains to determine $\delta \zeta$. An efficient method to 
deduce this was outlined in \cite{52} but followed a different but equivalent
strategy to that given in \cite{51}. In \cite{52} the divergence structure of 
the correlation function of $J$ or equivalently that of ${\cal O}$ was 
considered in the massless theory. This Green's function is indicated 
graphically in Figure \ref{figcorj}. Unlike say the correlation function of the
field strength all Feynman graphs contributing to this Green's function will 
only have two gluons emanating from each operator insertion. Strictly the LCO 
method will produce a non-zero mass for the gluon and therefore one should in 
principle determine the Green's function with massive gluon propagators. 
However as only the ultraviolet divergences are required for the
counterterm $\delta \zeta$ massless gluon propagators will suffice. The 
advantage of this, as noted in \cite{52}, was that the Green's function could 
be computed using the automatic Feynman diagram package {\sc Mincer},
\cite{69,70}. In following the same strategy here we use instead the recent 
extension of that programme which is {\sc Forcer}, \cite{53,54}. Both packages 
determine the divergences of massless $2$-point functions as a function of 
$\epsilon$. The {\sc Mincer} package was designed for three loop computations 
but with the current need for more precision in quantum field theory 
{\sc Forcer} was developed to calculate four loop graphs. This is a key point 
since the LCO formalism requires that $\delta \zeta$ has to be calculated to 
four rather than three loops here. For the two loop effective potential of 
\cite{51} the three loop term of $\delta \zeta$ was needed which is why 
{\sc Mincer} was the appropriate tool. Therefore we have used {\sc Forcer} to 
re-evaluate and verify the three loop counterterm before extending it to four 
loops here. As background there are $1$, $5$, $67$ and $1242$ Feynman graphs at
the respective succussive loop orders from one to four. These were generated by 
the {\sc Qgraf} package, \cite{71}, which is the starting point for the 
automatic computation of the Feynman graphs contributing to the four loop $J$ 
correlation function. Throughout we employed the symbolic manipulation language
{\sc Form}, \cite{72,73}, for handling the tedious amounts of algebra. Indeed 
the {\sc Forcer} package is written in {\sc Form}. Equally we made use of the 
{\sc Form} based {\tt color.h} module, \cite{72}, to evaluate the group theory
associated with each graph. That routine is based on the comprehensive article 
\cite{74} on non-abelian groups.

{\begin{figure}[ht]
\begin{center}
\includegraphics[width=13.0cm,height=8.0cm]{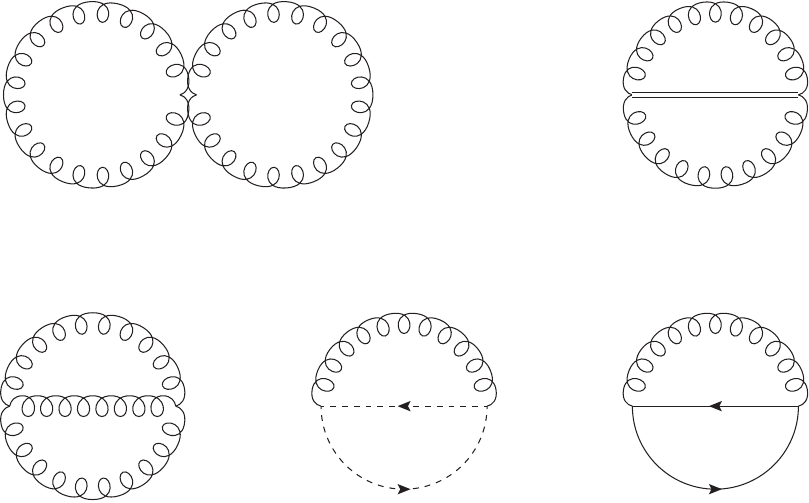}
\end{center}
\caption{Two loop graphs contributing to the effective potential.}
\label{figef2}
\end{figure}}

Consequently we have been able to deduce the counterterm $\delta \zeta$ to four
loops for an arbitrary colour group. As the full expression for it is large it 
has been provided in Appendix A with electronic versions of all the major 
expressions computed in this article provided in the attached data file. This 
means that the various quantities that are derived from $\delta \zeta$ are as 
large and also recorded in that appendix. However in order to represent the
procedure to derive the effective potential we will present the key equations
in this section for the case of $SU(3)$ and $\Nf$~$=$~$3$. Therefore the 
counterterm for the renormalization of $J$ is 
\begin{eqnarray}
\left. \frac{}{} \delta \zeta \right|^{SU(3)}_{\Nf=3} &=&
-~ \frac{12}{\epsilon} ~+~ 
3 \left[ \frac{27}{\epsilon^2} - \frac{82}{\epsilon} \right] a ~+~ 
\left[ - \frac{7290}{\epsilon^3} + \frac{26934}{\epsilon^2} 
+ [ 1443 \zeta_3 - 55109 ] \frac{1}{\epsilon} \right] \frac{a^2}{12}
\nonumber \\
&&
+~ \left[ \frac{5511240}{\epsilon^4} - \frac{24184980}{\epsilon^3}
- [ 801252 \zeta_3 + 56625408 ] \frac{1}{\epsilon^2}
\right. \nonumber \\
&& \left. ~~~~~
+ [ 15906120 \zeta_5 - 1781190 \zeta_4 - 1771416 \zeta_3 - 109560851 ]
\frac{1}{\epsilon} \right] \frac{a^3}{1152} \nonumber \\
&& +~ O(a^4)
\label{deltazsu3}
\end{eqnarray}
where $\zeta_n$ is the Riemann zeta function. Converting this to a 
renormalization group function via (\ref{deltadef}) leads to
\begin{eqnarray}
\left. \frac{}{} \delta(a) \right|^{SU(3)}_{\Nf=3} &=&
-~ 24 ~-~ 984 a ~+~ [ 1443 \zeta_3 - 55109 ] \frac{a^2}{2}
\nonumber \\
&&
+~ [ 15906120 \zeta_5 - 1781190 \zeta_4 - 1771416 \zeta_3 - 109560851 ]
\frac{a^3}{144} ~+~ O(a^4) ~. ~~~~
\end{eqnarray}
It is a simple matter now to perturbatively solve the consistency equation
that determines $\zeta(a)$ and find
\begin{eqnarray}
\left. \frac{}{} \frac{1}{\zeta(a)} \right|^{SU(3)}_{\Nf=3} &=&
\frac{3}{8} a ~-~ \frac{1189}{288} a^2 ~+~ 
[ 567891 \zeta_3 - 2627074 ] \frac{a^3}{155520}
\nonumber \\
&&
+\, [
72516983400 \zeta_5
- 9659825910 \zeta_4
- 22815926343 \zeta_3
- 80343163558 ] \frac{a^4}{235146240} \nonumber \\
&& +~ O(a^5) ~.
\label{zetasu3}
\end{eqnarray}
The final quantity required prior to computing the effective potential is the
renormalization constant
\begin{eqnarray}
\left. \frac{}{} Z_\zeta^{-1} \right|^{SU(3)}_{\Nf=3} &=&
1 ~+~ \frac{9 a}{2\epsilon} ~+~ 
\left[ - \frac{243}{\epsilon^2} + \frac{1025}{\epsilon} \right] \frac{a^2}{24} 
\nonumber \\
&&
+~ \left[ \frac{590490}{\epsilon^3} - \frac{1592865}{\epsilon^2} 
+ [ 6529841 - 16524 \zeta_3 ] \frac{1}{\epsilon} \right] 
\frac{a^3}{12960} ~+~ O(a^4) 
\label{Zzetasu3}
\end{eqnarray}
deduced from (\ref{deltazsu3}) and (\ref{zetasu3}) using (\ref{Zzdef}).

{\begin{figure}[ht]
\begin{center}
\includegraphics[width=15.0cm,height=12.0cm]{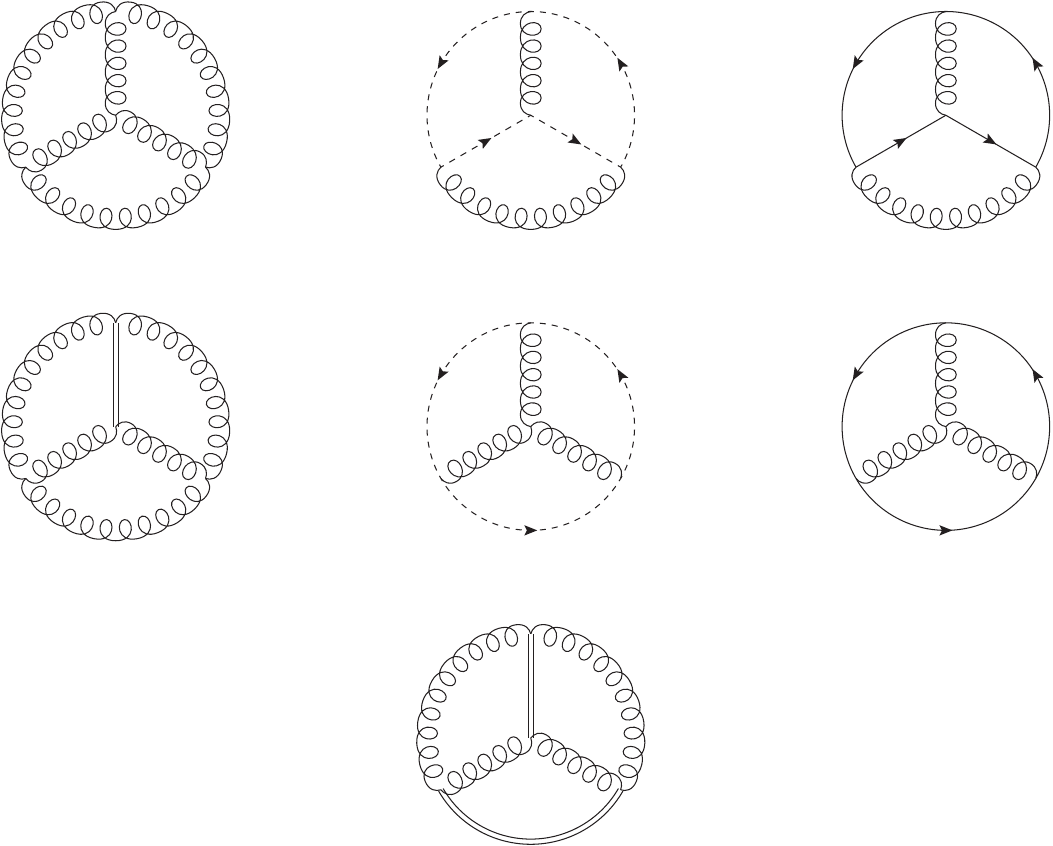}
\end{center}
\caption{Three loop benz graphs contributing to the effective potential.}
\label{figef3a}
\end{figure}}

The foundation has now been put in place from the LCO construction for the
effective potential of ${\cal O}$ to be determined. By this we mean the
renormalization constants of the variables have been found to the necessary
order to produce a finite three loop expression. As we will be using 
(\ref{wrgelin}) and (\ref{lagsig}) as the starting point for the process, since
for instance there is a linear source term for $\sigma$, we follow the standard
procedure for deducing an effective potential. At the outset it is worth noting
related articles in this area that were useful for this task. These were the 
three loop effective potentials given in \cite{75} for $O(N)$ $\phi^4$ theory, 
the application of the LCO method to a scalar theory, \cite{76}, and most 
usefully the comprehensive study of \cite{77} for a general renormalizable 
theory that includes the Standard Model. First, to set the scene we recall that
the one loop effective potential $V_1(\sigma)$ is deduced from an infinite set 
of $n$-point functions with a constant external $\sigma$ field. Summing this 
class of functions produces 
\begin{equation}
V_1(\sigma) ~=~ \frac{\sigma^2}{2g^2 \zeta(a) Z_\zeta} ~+~ \frac{(d-1)N_A}{2}
\int \frac{d^d k}{(2\pi)^d} \, 
\ln \left( k^2 ~+~ \frac{Z_m\sigma}{gZ_\zeta\zeta(a)} \right)
\end{equation}
where we have included the renormalization constants as these are necessary for
the higher order corrections. The factor of $(d-1)$ arises from the contraction
of the Lorentz indices of the massive Landau gauge gluon propagator with the 
mass given by, \cite{51,52}, 
\begin{equation}
m^2 ~=~ \frac{Z_m\sigma}{gZ_\zeta \zeta(a)}
\label{masseffbare}
\end{equation}
and $\NA$ appears from the trace of the adjoint colour indices.

To proceed to the two and three loop corrections one does not have to tediously
resum corrections to the leading $n$-point functions with a constant $\sigma$ 
field. Instead one equivalently computes the respective loop order vacuum 
graphs based on the interactions in $L^\sigma$ but with a new field 
$\tilde{\sigma}$. This is introduced by the simple shift, \cite{51,52},
\begin{equation}
\sigma ~=~ \langle \sigma \rangle ~+~ \tilde{\sigma}
\label{sigshft}
\end{equation}
where the expectation value of $\sigma$ is a constant. We note that the 
propagator for $\tilde{\sigma}$ derives from the coefficient of 
$\tilde{\sigma}^2$ in $L^\sigma$ after applying (\ref{sigshft}). Again we have 
used {\sc Qgraf} to generate the two and three loop vacuum graphs. The graphs
that contribute at two loops are shown in Figure \ref{figef2} where the spring 
lines are gluons, the dotted directed lines are quarks, the solid directed 
lines are Faddeev-Popov ghosts and the double solid lines correspond to 
$\tilde{\sigma}$. We note that the interactions of the latter involve 
$\zeta(a)$. At three loops more graphs contribute to the effective potential 
and these are shown across three Figures. The ones based on the benz topology 
are given in Figure \ref{figef3a} while those based on the ladder topology 
given in Figure \ref{figef3b}. The remaining graphs are provided in Figure 
\ref{figef3c} which include the one loop propagator corrections to the two loop
graphs of Figure \ref{figef2} from massive gluon snail graphs as well as vertex
corrections to the same graphs.

The next stage in the process is the evaluation of the $5$ two loop and $29$
three loop vacuum graphs for $V(\sigma)$. As with the derivation of 
$\delta \zeta$ at four loops we have carried this out by an automatic Feynman
diagram computation. Unlike that case the subsequent integrals are massive
since we assume $\langle \sigma \rangle$ is non-zero and therefore we cannot 
apply the {\sc Forcer} algorithm. Instead we employ the Laporta algorithm, 
\cite{45}, where all the integrals contributing to a vacuum graph are reduced 
to a small set of core master integrals. In practical terms we used the 
{\sc Reduze} implementation of the algorithm, \cite{78}, to achieve this. While
analytic expressions for three (and higher) loop vacuum master graphs have been
determined by various methods we use the results provided in \cite{79} for our 
automatic computation. This is because \cite{79} gathers and summarizes the 
earlier work of \cite{80} on the vacuum benz topologies as well as that of 
others \cite{81,82,83} with the same conventions. To ensure that the final 
result is ultimately finite, as the counterterms are already determined, the 
master two loop master integrals have to be expanded to the requisite order in 
$\epsilon$. Such terms are available in the expressions recorded in \cite{79}. 
In addition $V_1(\sigma)$ has to be expanded to $O(\epsilon^2)$ prior to the 
substitution of the various renormalization constants.

{\begin{figure}[ht]
\begin{center}
\includegraphics[width=14.0cm,height=15.0cm]{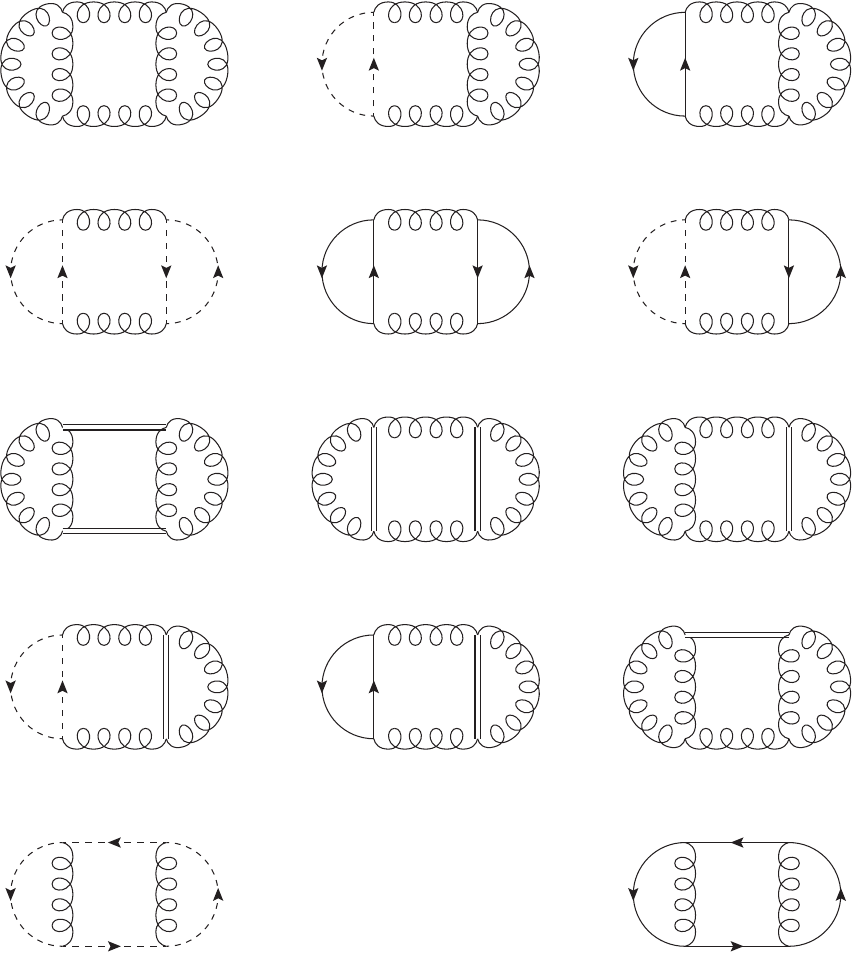}
\end{center}
\caption{Three loop ladder graphs contributing to the effective potential.}
\label{figef3b}
\end{figure}}

Collecting the various contributions from the graphs together with the
renormalization constants determined to the necessary order we arrive at the
three loop effective potential in the $\MSbar$ scheme. As our focus in this 
section is for $SU(3)$ and $\Nf$~$=$~$3$ in $SU(3)$ we note
\begin{eqnarray}
\left. \frac{}{} V(\sigma) \right|^{SU(3)}_{\Nf=3} &=& 
\left[
\frac{4}{3}
+ \left[
6 \overline{\ln}(\bar{g} \bar{\sigma})
- \frac{1594}{81}
\right] a \right. \nonumber \\
&& \left.
+ \left[ 
\frac{779}{60} \zeta_3
+ \frac{45}{2} \zeta_2
- \frac{3340906}{10935}
+ \frac{297}{2} \sqrt{3} \mbox{Ls}_2(\twopithree)
+ \frac{1106}{9} \overline{\ln}(\bar{g} \bar{\sigma}) 
- \frac{27}{2} \overline{\ln}^2(\bar{g} \bar{\sigma}) 
\right] a^2
\right. \nonumber \\
&& \left.
+ \left[ 
\frac{828955}{756} \zeta_5
+ \frac{69153}{64} \zeta_4
+ \frac{7742747}{12960} \zeta_3
+ \frac{40319}{96} \zeta_2
- \frac{91909989967}{9447840}
\right. \right. \nonumber \\
&& \left. \left. ~~~~
- 1769 \mbox{Li}_4(\half)
- \frac{3321}{32} \mbox{Ls}^2_2(\twopithree)
+ \frac{1769}{4} \ln^2(2) \zeta_2
- \frac{1769}{24} \ln^4(2)
\right. \right. \nonumber \\
&& \left. \left. ~~~~
+ \frac{121}{6} \sqrt{3} \pi^3
+ \frac{5659}{4} \sqrt{3} \mbox{Ls}_2(\twopithree)
\right. \right. \nonumber \\
&& \left. \left. ~~~~
+ \left[
\frac{1924658}{405}
- \frac{2541}{40} \zeta_3
- \frac{6075}{16} \zeta_2
- \frac{8019}{4} \sqrt{3} \mbox{Ls}_2(\twopithree)
\right]
\overline{\ln}(\bar{g} \bar{\sigma})
\right. \right. \nonumber \\
&& \left. \left. ~~~~
- \frac{1767}{2} \overline{\ln}^2(\bar{g} \bar{\sigma}) 
+ \frac{243}{4} \overline{\ln}^3(\bar{g} \bar{\sigma}) 
\right] a^3 \right] \bar{\sigma}^2 ~+~ O(a^4)
\label{potsu3}
\end{eqnarray}
where $\bar{g}$~$=$~$\frac{g}{4\pi}$, 
$\mbox{Li}_n(z)$ is the polylogarithm function,
\begin{equation}
\overline{\ln}(z) ~=~ \ln \left( \frac{z}{\mu^2} \right) ~,
\end{equation}
the log-sine function is 
\begin{equation}
\mbox{Ls}_n(\theta) ~=~ -~ \int_0^{\theta} dz \, \ln^{n-1} 
\left| 2 \sin (\half z) \right| 
\end{equation} 
noting that $\mbox{Ls}_2(\theta)$~$=$~$\mbox{Cl}_2(\theta)$ with 
$\mbox{Cl}_2(\theta)$ denoting the Clausen function and we have introduced the 
shorthand notation
\begin{equation}
\sigma ~=~ \frac{9 N_A}{(13 C_A - 8 T_F \Nf)} \bar{\sigma} 
\end{equation}
with the colour dependent factor arising from the coefficient of the leading
term of $\zeta(a)$. The full colour group expression for $V(\sigma)$ is given
in Appendix A. In arriving at that we have reproduced the two loop Yang-Mills
theory result of \cite{51} and the subsequent extension to $\Nf$ massless 
quarks, \cite{52}. The potential is as one would expect in that the three loop 
term is cubic in $\overline{\ln} (\bar{g}\bar{\sigma})$.

{\begin{figure}[ht]
\begin{center}
\includegraphics[width=14.0cm,height=7.25cm]{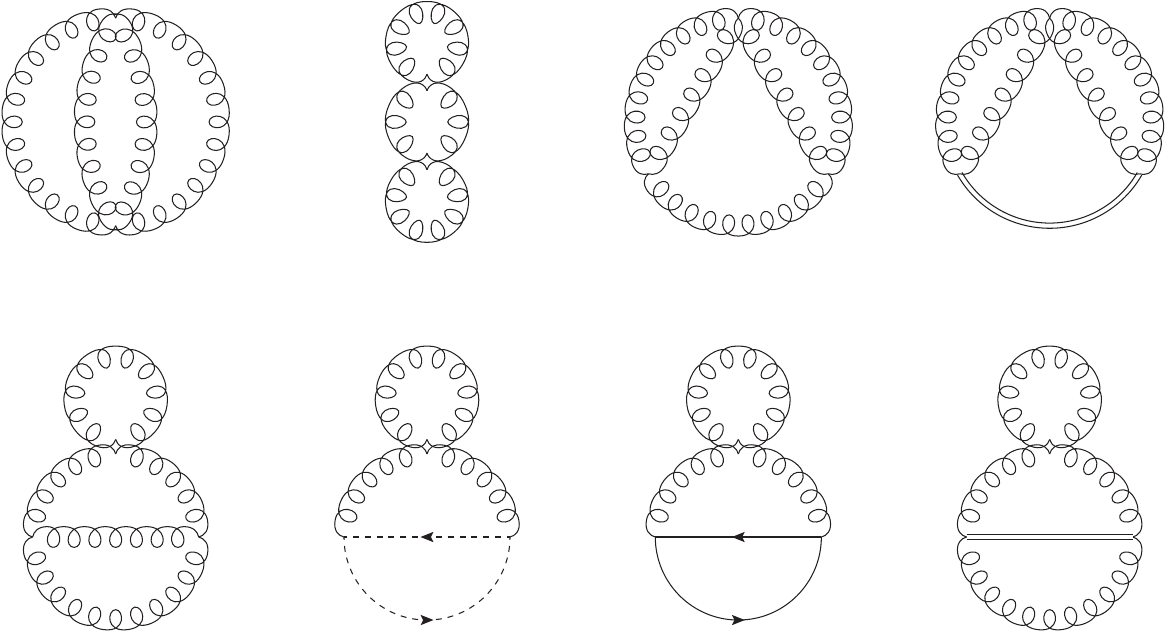}
\end{center}
\caption{Remaining three loop graphs contributing to the effective potential.}
\label{figef3c}
\end{figure}}

\sect{Analysis.}

{\begin{table}[ht]
\begin{center}
\begin{tabular}{|c||c|c|c|}
\hline
$\Nf$ & $a^{(1)}$ & $a^{(2)}$ & $a^{(3)}$ \\
\hline
$\!\!0$ & $0.064171$ & $0.046493$ & $0.035626$ \\
$2$ & $0.073973$ & $0.061090$ & $0.044562$ \\
$3$ & $0.079941$ & $0.074190$ & $0.051766$ \\
$4$ & $0.086787$ & $0.100128$ & $0.063391$ \\
$5$ & $0.094641$ & --------- & $0.090027$ \\
\hline
\end{tabular}
\end{center}
\begin{center}
\caption{Values of $SU(3)$ coupling constants at the minimum of the potential
for $\Nf$ massless quarks.}
\label{ccsu3}
\end{center}
\end{table}}

{\begin{table}[hb]
\begin{center}
\begin{tabular}{|c||c|c|c|}
\hline
$\Nf$ & $a^{(1)}$ & $a^{(2)}$ & $a^{(3)}$ \\
\hline
$0\!\!$ & $0.096257$ & $0.069739$ & $0.053594$ \\
$2$ & $0.120805$ & $0.111985$ & $0.078461$ \\
$3$ & $0.137908$ & $0.217124$ & $0.113177$ \\
$4$ & $0.138742$ & --------- & --------- \\
$5$ & $0.146341$ & --------- & --------- \\
\hline
\end{tabular}
\end{center}
\begin{center}
\caption{Values of $SU(2)$ coupling constants at the minimum of the potential
for $\Nf$ massless quarks.}
\label{ccsu2}
\end{center}
\end{table}}

We are now in a position to refine the two loop estimates for 
$\langle \frac{1}{2} { A_\mu^a }^2 \rangle$ given in \cite{51,52} and from them
to deduce a value for the gluon mass. The analysis in this section will focus 
again on $SU(3)$ with $\Nf$~$=$~$3$ in order to illustrate the method but we 
will record results for the gluon mass for other values of $\Nf$ as well as for
$SU(2)$. First, to assist with this we record the numerical expression for the 
effective potential which is
\begin{eqnarray}
\left. \frac{}{} V(\bar{\sigma}) \right|^{SU(3)}_{\Nf=3} &=& 
\left[ 1.333333 
+ \left[ 6.000000 \overline{\ln} (\bar{g} \bar{\sigma}) - 19.679012 \right] a  
\right. \nonumber \\
&& \left. 
+~ \left[ - 13.500000 \overline{\ln}^2 (\bar{g} \bar{\sigma})  
+ 122.888889 \overline{\ln} (\bar{g} \bar{\sigma}) 
- 78.871262 \right] a^2                                    
\right. \nonumber \\
&& \left. 
+~ \left[ 60.750000 \overline{\ln}^3 (\bar{g} \bar{\sigma}) 
- 883.500000 \overline{\ln}^2 (\bar{g} \bar{\sigma}) 
+ 1701.846389 \overline{\ln} (\bar{g} \bar{\sigma}) 
\right. \right. \nonumber \\
&& \left. \left. ~~~~~
- 3902.054942 \right] a^3
\right] \bar{\sigma}^2 ~+~ O(a^4) ~.
\label{effpotsu3num}
\end{eqnarray}
The stationary points are deduced by solving 
\begin{equation}
\frac{d V}{d \bar{\sigma}} ~=~ 0 
\end{equation}
and from (\ref{effpotsu3num}) we have 
\begin{eqnarray}
\left.  \frac{d V}{d \bar{\sigma}} \right|^{SU(3)}_{\Nf=3} &=& 
\left[ 2.666667
+ \left[ 12.000000 \overline{\ln} (\bar{g} \bar{\sigma}) - 33.358025 \right] a
\right. \nonumber \\
&& \left. 
+~ \left[ - 27.000000 \overline{\ln}^2(\bar{g} \bar{\sigma})
+ 218.777778 \overline{\ln} (\bar{g} \bar{\sigma})
- 34.853636 \right]  a^2
\right. \nonumber \\
&& \left. 
+~ \left[ 121.500000 \overline{\ln}^3 (\bar{g} \bar{\sigma})
- 1584.750000 \overline{\ln}^2 (\bar{g} \bar{\sigma})
+ 1636.692777 \overline{\ln} (\bar{g} \bar{\sigma})
\right. \right. \nonumber \\
&& \left. \left. ~~~~~
- 6102.263495 \right] a^3
\right] \bar{\sigma} ~+~ O(a^4) ~.
\end{eqnarray}
In order to extract the location of the stationary point we follow the 
prescription given in \cite{46,47,51} and set the scale to be where the 
logarithm vanishes which is 
\begin{equation}
\bar{g} \bar{\sigma} ~=~ \mu^2 
\label{mincond}
\end{equation}
which produces several solutions. There is one stationary point at 
$\bar{\sigma}$~$=$~$0$ which corresponds to the usual perturbative vacuum with 
$V(0)$~$=$~$0$. At one and higher loops this is a local maximum but in the 
absence of loop corrections it a minimum. Non-zero solutions for $\bar{\sigma}$
result from the remaining cubic in $a$ and these correspond to the global 
minimum of the potential. We have solved for the coupling constant values 
numerically and the $SU(3)$ results are given in Table \ref{ccsu3} with 
$a^{(L)}$ indicating the solution at $L$ loops. In \cite{51,52} a different 
convention was used for the coupling constant which was denoted by $y$. It is 
related to $a$ by factor of $\Nc$. Here we prefer to use $a$ as the variable 
since the rescaling by $\Nc$ or in effect $C_A$ is arbitrary and ineffectual. 
Moreover additional rank $4$ Casimirs appear in the effective potential at 
three loops as is evident in (\ref{effpotarb}).

{\begin{table}[ht]
\begin{center}
\begin{tabular}{|c||c|c|c|}
\hline
$\Nf$ & $L$~$=$~$1$ & $L$~$=$~$2$ & $L$~$=$~$3$ \\
\hline
$0\!\!$ & $2.030604$ & $2.003773$ & $2.591158$ \\
$2$ & $2.012200$ & $1.878750$ & $2.418769$ \\
$3$ & $2.003627$ & $1.802659$ & $2.297279$ \\
$4$ & $1.996403$ & $1.702877$ & $2.120771$ \\
$5$ & $1.991927$ & --------- & $1.806990$ \\
\hline
\end{tabular}
\end{center}
\begin{center}
\caption{Values of 
$\bar{g} \langle \bar{\sigma} \rangle/\Lambda_{\MSbars}$ for $SU(3)$ at $L$ 
loops using (\ref{effccmass}).}
\label{vevsu3cc}
\end{center}
\end{table}}

{\begin{table}[hb]
\begin{center}
\begin{tabular}{|c||c|c|c|}
\hline
$\Nf$ & $L$~$=$~$1$ & $L$~$=$~$2$ & $L$~$=$~$3$ \\
\hline
$0\!\!$ & $2.030604$ & $2.124240$ & $2.654008$ \\
$2$ & $2.012200$ & $1.986479$ & $2.467973$ \\
$3$ & $2.003627$ & $1.890028$ & $2.334335$ \\
$4$ & $1.996403$ & $1.744687$ & $2.142946$ \\
$5$ & $1.991927$ & --------- & $1.826447$ \\
\hline
\end{tabular}
\end{center}
\begin{center}
\caption{Values of 
$\bar{g} \langle \bar{\sigma} \rangle/\Lambda_{\MSbars}$ for $SU(3)$ at $L$ 
loops using (\ref{vevmassdef}).}
\label{vevsu3}
\end{center}
\end{table}}

In Table \ref{ccsu3} there is no solution for the coupling at the two loop
stationary point for $\Nf$~$=$~$5$ and the $SU(3)$ group since the quadratic 
equation in $a$ produces complex conjugate roots. For the case of $SU(2)$ the 
situation is similar but arises at a lower value of $\Nf$ with those stationary
point couplings given in Table \ref{ccsu2}. While both tables indicate that the
non-zero stationary point coupling decreases with loop order what is more 
important is the behaviour of the effective gluon mass induced by the non-zero 
value of $\langle \bar{\sigma} \rangle$. Therefore we have to translate the 
stationary point couplings to a value for the gluon mass. This will be a two 
stage process. First using the values found for $a$ at the stationary point of 
the three loop potential we obtain estimates for 
$\bar{g} \langle \bar{\sigma} \rangle$ using the relation (\ref{mincond})
before repeating the process provided in \cite{51,52}. This is based on the 
relation between the running coupling constant and a mass scale in units of the
$\Lambda$ parameter. As we have focussed on the $\MSbar$ scheme throughout our 
reference scale will therefore be $\Lambda_{\MSbars}$. Solving the first order 
differential equation that defines the $\beta$-function in a perturbative way 
allows us to either express the coupling constant as a function of mass which 
we denote by $a^{(L)}(m)$ at each loop order $L$ or
$\bar{g} \langle \bar{\sigma} \rangle$ as a function of the coupling constant. 
We will use $m^{(L)}(a)$ to denote a generic mass scale. For reference to three
loops the explicit expressions are 
\begin{eqnarray}
a^{(1)}(m) &=&
-~ \frac{1}{\beta_0 \ln \left( \frac{m^2}{\Lambda^2} \right)} \nonumber \\
a^{(2)}(m) &=&
-~ \frac{1}{\beta_0 \ln \left( \frac{m^2}{\Lambda^2} \right)} 
- \frac{\beta_1 \ln \left(\ln \left( \frac{m^2}{\Lambda^2} \right) \right)}
{\beta_0^3 \ln^2 \left( \frac{m^2}{\Lambda^2} \right)} \nonumber \\
a^{(3)}(m) &=&
-~ \frac{1}{\beta_0 \ln \left( \frac{m^2}{\Lambda^2} \right)}
- \frac{\beta_1 \ln \left(\ln \left( \frac{m^2}{\Lambda^2} \right) \right)}
{\beta_0^3 \ln^2 \left( \frac{m^2}{\Lambda^2} \right)}
\nonumber \\
&&
-~ \left[ \beta_1^2 
\left[ \ln^2 \left(\ln \left( \frac{m^2}{\Lambda^2} \right) \right) 
- \ln \left( \ln \left( \frac{m^2}{\Lambda^2} \right) \right) - 1 \right]
+ \beta_0 \beta_2 \right] 
\frac{1}{\beta_0^5 \ln^3 \left( \frac{m^2}{\Lambda^2} \right)} 
\label{effccmass}
\end{eqnarray}
and
\begin{eqnarray}
m^{(1)}(a) &=& \Lambda
\exp \left[ -~ \frac{1}{2\beta_0 a} \right] \nonumber \\
m^{(2)}(a) &=& \Lambda
\exp \left[ -~ \frac{1}{2\beta_0 a} ~-~
\frac{\beta_1}{2\beta_0^2} \ln ( - \beta_0 a ) \right] \nonumber \\
m^{(3)}(a) &=& \Lambda
\exp \left[ -~ \frac{1}{2\beta_0 a} ~-~
\frac{\beta_1}{2\beta_0^2} \ln ( - \beta_0 a ) ~+~
\frac{1}{2\beta_0^3} \left[ \beta_1^2 - \beta_0 \beta_2 \right] a \right] ~.
\label{vevmassdef}
\end{eqnarray}
for any renormalization scheme. Specifically the $\beta$-function coefficients 
are, \cite{84,85,86,87,88,89},
\begin{eqnarray}
\beta_0 &=& -~ \left[ \frac{11}{3} C_A - \frac{4}{3} T_F \Nf \right]
\nonumber \\
\beta_1 &=& -~ \left[ \frac{34}{3} C_A^2 - 4 T_F C_F \Nf 
- \frac{20}{3} T_F \Nf C_A \right] \nonumber \\
\beta_2 &=& -~ \left[ \frac{2857}{54} C_A^3 + 2 C_F^2 T_F \Nf 
- \frac{205}{9} C_F C_A T_F \Nf - \frac{1415}{27} C_A^2 T_F \Nf 
\right. \nonumber \\
&& \left. ~~~~
+ \frac{44}{9} C_F T_F^2 \Nf^2
+ \frac{158}{27} C_A T_F^2 \Nf^2 \right] 
\label{betadef}
\end{eqnarray}
for reference. Using the values for $a^{(L)}$ given in Tables \ref{ccsu3} and 
\ref{ccsu2} it is a straightforward exercise to substitute them into the right 
side of (\ref{vevmassdef}). To deduce estimates for 
$\bar{g} \langle \bar{\sigma} \rangle$ from (\ref{effccmass}) requires more 
work in that we searched for a value of $\frac{m}{\Lambda}$ that gave the 
respective values of $a^{(L)}$ in each of these tables. One reason for using 
two methods rests in the fact that we are working with a truncation by loop 
order. If the all orders expressions for (\ref{effccmass}) and 
(\ref{vevmassdef}) were both available then each would be the precise inverse 
function of the other. So the solution to either would give the same value for 
the mass. In the truncated case the discrepancy between them would be a rough 
measure of the accuracy at each loop order. The results of this analysis for 
both approaches are recorded in Tables \ref{vevsu3cc} and \ref{vevsu3} for 
$SU(3)$ and the respective $SU(2)$ estimates are provided in Tables 
\ref{vevsu2cc} and \ref{vevsu2}. Several initial comments on Tables 
\ref{vevsu3cc} to \ref{vevsu2} are apt. First the estimates up to $\Nf$~$=$~$3$
at one and two loops in Tables \ref{vevsu3} and \ref{vevsu2} are in agreement 
with \cite{51,52} while the remainder are new. We have chosen to include higher
values of $\Nf$ to illustrate that at a certain point, which differs for each 
group, there is not always a solution for the coupling at a stationary point. 
Next the estimates for $\bar{g} \langle \bar{\sigma} \rangle$ at one loop in 
the pair of tables for each group are identical. This follows trivially from 
the fact that $a^{(1)}(m)$ and $m^{(1)}(a)$ are formal inverses of each other 
which is evident from (\ref{effccmass}) and (\ref{vevmassdef}). 

{\begin{table}[ht]
\begin{center}
\begin{tabular}{|c||c|c|c|}
\hline
$\Nf$ & $L$~$=$~$1$ & $L$~$=$~$2$ & $L$~$=$~$3$ \\
\hline
$0\!\!$ & $2.030604$ & $2.003773$ & $2.585258$ \\
$2$ & $1.993346$ & $1.793178$ & $2.269245$ \\
$3$ & $1.973484$ & $1.624988$ & $1.950830$ \\
$4$ & $2.164616$ & --------- & --------- \\
$5$ & $2.349416$ & --------- & --------- \\
\hline
\end{tabular}
\end{center}
\begin{center}
\caption{Values of $\bar{g} \langle \bar{\sigma} \rangle/\Lambda_{\MSbars}$ for
$SU(2)$ at $L$ loops using (\ref{effccmass}).}
\label{vevsu2cc}
\end{center}
\end{table}}

{\begin{table}[ht]
\begin{center}
\begin{tabular}{|c||c|c|c|}
\hline
$\Nf$ & $L$~$=$~$1$ & $L$~$=$~$2$ & $L$~$=$~$3$ \\
\hline
$0\!\!$ & $2.030604$ & $2.124240$ & $2.647960$ \\
$2$ & $1.993346$ & $1.882611$ & $2.306476$ \\
$3$ & $1.973484$ & $1.584461$ & $1.969801$ \\
$4$ & $2.164616$ & --------- & --------- \\
$5$ & $2.349416$ & --------- & --------- \\
\hline
\end{tabular}
\end{center}
\begin{center}
\caption{Values of $\bar{g} \langle \bar{\sigma} \rangle/\Lambda_{\MSbars}$ for
$SU(2)$ at $L$ loops using (\ref{vevmassdef}).}
\label{vevsu2}
\end{center}
\end{table}}

{\begin{table}[ht]
\begin{center}
\begin{tabular}{|c||c|c|}
\hline
$\Nf$ & $\Lambda_{\MSbars}$ & Reference \\
\hline
\rule{0pt}{12pt}
$\!\!0$ & $224^{+8}_{-5}$ & \cite{90} \\
$2$ & $294^{+10}_{-10}$ & \cite{91} \\
$3$ & $339^{+10}_{-10}$ & \cite{92} \\
$4$ & $296^{+10}_{-10}$ & \cite{92} \\
$5$ & $213^{+8}_{-8}$ & \cite{92} \\
\hline
\end{tabular}
\end{center}
\begin{center}
\caption{Values of $\Lambda_{\MSbars}$ in MeV.}
\label{lambdaval}
\end{center}
\end{table}}

The second stage in the process to estimate the gluon mass is to translate the
values of $\bar{g} \langle \bar{\sigma} \rangle$ into the effective mass which
is defined by 
\begin{equation}
m^2_{\mbox{\footnotesize{eff}}} ~=~ 
\frac{\langle \bar{\sigma} \rangle}{g \zeta(a)}
\label{masseffdef}
\end{equation}
based on (\ref{masseffbare}). This definition differs from that used in 
\cite{46,47,51,52} which took the value of 
$\bar{g} \langle \bar{\sigma} \rangle$ for the gluon mass. If one were to use
that approach here it is evident that there would be convergence concerns given
the jump in the three loop values in comparison with the two loop ones from 
examining Tables \ref{vevsu3cc} to \ref{vevsu2} for both groups. In light of 
this we have taken (\ref{masseffdef}) as a more appropriate quantity to use. 
Consequently we have to compute a value for $\zeta(a)$ for each group and $\Nf$
value which is achieved by evaluating $\zeta(a)$ using the coupling constants 
given in Tables \ref{ccsu3} and \ref{ccsu2}. However naively substituting those
numbers into the perturbative expression for $1/(a\zeta(a))$, either
(\ref{zetasu3}) or (\ref{zetagen}), will lead to convergence issues. Therefore 
to deduce each adjustment factor we have calculated Pad\'{e} approxminants to 
$1/(a\zeta(a))$ using the one, two and three loop terms of (\ref{zetasu3}) and 
(\ref{zetagen}) to find the numbers to multiply each entry in Tables 
\ref{vevsu3cc} to \ref{vevsu2}. In particular we have used the $[0,1]$, $[0,2]$
and $[1,2]$ Pad\'{e} approximants at the respective orders. For the $SU(3)$ 
case we can also include the values of $\Lambda_{\MSbars}$ which are summarized
in Table \ref{lambdaval}. Therefore combining the three values of $\bar{g} 
\bar{\sigma}$ and $1/(a\zeta(a))$ we arrive at the values given in Tables 
\ref{masssu3cc} and \ref{masssu3}. The errors are derived from those given in 
Table \ref{lambdaval}.

{\begin{table}[ht]
\begin{center}
\begin{tabular}{|c||c|c|c|}
\hline
$\Nf$ & $L$~$=$~$1$ & $L$~$=$~$2$ & $L$~$=$~$3$ \\
\hline
\rule{0pt}{12pt}
$\!\!0$ & $333^{+12}_{-6}$ & $303^{+11}_{-7}$ & $319^{+12}_{-7}$ \\
$2$ & $432^{+15}_{-14}$ & $354^{+12}_{-12}$ & $355^{+12}_{-12}$ \\
$3$ & $495^{+15}_{-15}$ & $374^{+11}_{-11}$ & $369^{+10}_{-11}$ \\
$4$ & $430^{+14}_{-15}$ & $280^{+9~}_{-9}$ & $287^{+10}_{-9}$ \\
$5$ & $308^{+11}_{-12}$ & -------- & $166^{+6}_{-7}$ \\
\hline
\end{tabular}
\end{center}
\begin{center}
\caption{Values of $m_{\mbox{\footnotesize{eff}}}$ in MeV for $SU(3)$ at $L$ 
loops based on Table \ref{vevsu3cc}.}
\label{masssu3cc}
\end{center}
\end{table}}

{\begin{table}[ht]
\begin{center}
\begin{tabular}{|c||c|c|c|}
\hline
$\Nf$ & $L$~$=$~$1$ & $L$~$=$~$2$ & $L$~$=$~$3$ \\
\hline
\rule{0pt}{12pt}
$\!\!0$ & $333^{+12}_{-6}$ & $321^{+12}_{-7}$ & $327^{+12}_{-7}$ \\
$2$ & $432^{+15}_{-14}$ & $374^{+13}_{-13}$ & $361^{+13}_{-11}$ \\
$3$ & $495^{+15}_{-15}$ & $392^{+14}_{-11}$ & $375^{+11}_{-12}$ \\
$4$ & $430^{+14}_{-15}$ & $287^{+12}_{-10}$ & $290^{+10}_{-9}$ \\
$5$ & $308^{+11}_{-12}$ & --------- & $168^{+6}_{-7}$ \\
\hline
\end{tabular}
\end{center}
\begin{center}
\caption{Values of $m_{\mbox{\footnotesize{eff}}}$ in MeV for $SU(3)$ at $L$ 
loops based on Table \ref{vevsu3}.}
\label{masssu3}
\end{center}
\end{table}}

What is evident from both Tables \ref{masssu3cc} and \ref{masssu3} is that 
there is a degree of stability in the two and three loop effective gluon mass 
estimates up to $\Nf$~$=$~$4$. This is probably not unrelated to the coupling
constants at these orders being smaller than the one loop value for
$\Nf$~$\neq$~$0$ and therefore more within the range of perturbative 
reliability. Another feature is that the mass increases up to $\Nf$~$=$~$3$. 
This behaviour though needs to be tempered by recalling that we have assumed 
the quarks are massless. The inclusion of a quark mass would undoubtedly 
complicate the structure of the effective potential and thereby affect the 
effective gluon mass estimates. As we chose to gauge the truncation discrepancy
by extracting a value for $\bar{g} \langle \bar{\sigma} \rangle$ from the 
couplings in two different ways it is worth commenting on this. If one compares
say the $\Nf$~$=$~$0$ two loop effective mass values in Tables \ref{masssu3cc} 
and \ref{masssu3} there is roughly a $6\%$ difference. This reduces to around
$2.5\%$ at three loops. A similar narrowing is present for the other $\Nf$
cases. One main feature is that in the Yang-Mills theory case there is a marked
degree of consistency at each loop order. One way of providing a rough ballpark 
estimate of the effective gluon mass using the LCO approach is to average the
three loop value in the tables. For example for Yang-Mills theory this would 
give $m_{\mbox{\footnotesize{eff}}}$~$=$~$323$ MeV.

Other methods have produced gluon mass estimates which we mention for balance. 
For instance, the study of \cite{43} investigated the effect of including the 
Gribov mass, \cite{10}, in a model which also included a gluon mass operator. 
The main observation was that both mass scales were roughly equal and around 
$500$ MeV. In particular the two mass parameters appear in the gluon propagator
that can be derived from various modifications of Gribov's original action, 
\cite{10}. Zwanziger's reformulation of the Gribov action as a local 
renormalizable one, \cite{11,12,13,14,15,16,17,18,19,20,21}, provided the basis
for various extensions that included a gluon mass parameter such as
\cite{22,23,24}. While such a study produces mass that is of the accepted 
magnitude this was achieved by fitting to a classical propagator and therefore 
omitted quantum effects. This has been extended in later investigations. For
instance, one and two loop corrections to the gluon propagator have been 
calculated using the Landau gauge Lagrangian with a gluon mass term which is 
justified by the interpretation of the mass being related to Gribov copies 
\cite{35,36,37,38,39,40,41,42}. The subsequent gluon mass estimates were 
deduced by fitting the one and two loop propagators, that were calculated 
analytically, \cite{37,40,41,42}, against Landau gauge lattice data producing 
respective estimates of $350$ MeV and $330$ MeV. A more recent study, 
\cite{44}, using a functional renormalization group approach computed the 
effective potential of the field strength. From the value of the resulting 
field strength condenate an effective gluon mass gap of $0.312(27)$ GeV was 
found which was commensurate with the lattice value of $0.3536(11)$ GeV for the
mass gap derived from the lattice data of \cite{93}. These two values, as well 
as a similar gluon mass estimate, \cite{37,40,41,42}, extracted from different 
lattice data, indicates a relative degree of consistency from various 
directions. Therefore it is intersting that the LCO method produced a three 
loop gluon mass estimate that is very similar to these independent Lagrangian 
based methods. 

{\begin{table}[ht]
\begin{center}
\begin{tabular}{|c||c|c|c|}
\hline
$\Nf$ & $L$~$=$~$1$ & $L$~$=$~$2$ & $L$~$=$~$3$ \\
\hline
\rule{0pt}{12pt}
$\!\!0$ & $1.488526$ & $1.352525$ & $1.446185$ \\
$2$ & $1.454108$ & $1.098230$ & $1.128747$ \\
$3$ & $1.434460$ & $0.761928$ & $0.937949$ \\
\hline
\end{tabular}
\end{center}
\begin{center}
\caption{Values of $m_{\mbox{\footnotesize{eff}}}/\Lambda_{\MSbars}$ for 
$SU(2)$ at $L$ loops based on Table \ref{vevsu2cc}.}
\label{masssu2cc}
\end{center}
\end{table}}

{\begin{table}[ht]
\begin{center}
\begin{tabular}{|c||c|c|c|}
\hline
$\Nf$ & $L$~$=$~$1$ & $L$~$=$~$2$ & $L$~$=$~$3$ \\
\hline
\rule{0pt}{12pt}
$\!\!0$ & $1.488526$ & $1.433839$ & $1.481260$ \\
$2$ & $1.454108$ & $1.153003$ & $1.147266$ \\
$3$ & $1.434460$ & $0.742926$ & $0.946343$ \\
\hline
\end{tabular}
\end{center}
\begin{center}
\caption{Values of $m_{\mbox{\footnotesize{eff}}}/\Lambda_{\MSbars}$ for 
$SU(2)$ at $L$ loops based on Table \ref{vevsu2}.}
\label{masssu2}
\end{center}
\end{table}}

We complete our analysis by providing similar estimates for the $SU(2)$ 
effective gluon mass. The situation differs here in that we do not have values
for $\Lambda_{\MSbars}$ and so cannot give mass estimates in any physical 
units. Instead we have provided 
$m_{\mbox{\footnotesize{eff}}}/\Lambda_{\MSbars}$ for the three values of $\Nf$
where there were coupling constant solutions at each loop order. These are 
recorded in Tables \ref{masssu2cc} and \ref{masssu2} where the respective 
values from the Pad\'{e} approximants to $1/(a\zeta(a))$ have been included. A 
similar pattern to the $SU(3)$ case is apparent. There is a degree of stability
for Yang-Mills theory across loop orders for both tables. For $\Nf$~$=$~$2$ the
two and three loop estimates are similar but for $\Nf$~$=$~$3$ there is less 
consistency across loop order. This is parallel to the $\Nf$~$=$~$4$ and $5$ 
cases for $SU(3)$ since the latter is the value before which there are no two 
or three loop solutions for the minimum of the effective potential. The three 
loop effective gluon mass estimate in that situation appears to be out of line 
with the pattern of values for lower $\Nf$. This is perhaps an indicator of the
loss of a minimal effective potential solution for higher $\Nf$ and this 
appears to be borne out for the $SU(2)$ group as well.

\sect{Discussion.} 

We have applied the LCO formalism to a new loop order for any theory and in
particular a gauge theory. Our main focus has centred on QCD with $\Nf$ 
massless quarks. The key observation is that for Yang-Mills theory the three
loop corrections to the effective gluon mass derived from the minimum of the 
effective potential demonstrate a degree of convergence. For the $SU(3)$ colour
group this is centred roughly around the value of $323$ MeV for the effective
mass. This is not inconsistent, for instance, with values from other methods 
such as extracting a mass by fitting perturbative gluon propagators to lattice 
data or functional renormalization group techniques. The same analysis showed 
that this apparent convergence was evident for $SU(2)$ Yang-Mills theory as 
well. The situation for non-zero $\Nf$ was not as clear cut. This was primarily
because, as noted in \cite{52}, there was a marked difference between the one
and two loop values of the effective mass computed there albeit with a 
different definition from the one used here. With the current definition 
(\ref{masseffdef}) this is also apparent but interestingly Tables
\ref{masssu3cc} and \ref{masssu3} support the notion that the two and three
loop values for the effective mass have stabilized when further corrections are
included. Of course this is in the simplified scenario of massless quarks which
is not truly realistic. What would be needed is a generalization of the LCO 
method to induce quark masses from an extended effective potential. This does 
not seem straightforward if one naively examines the core formalism. The LCO 
construction for the gluon mass operator benefits from this having mass
dimension $2$ with an associated mass dimension $2$ source field that therefore
requires a quadratic source term on renormalizability grounds. Using the same 
dimensional analysis the quark mass operator would have dimension $3$ meaning 
its source field would be dimension $1$. Therefore aside from the mixing of the
two types of sources to produce a cubic term, quartic quark mass operator 
source terms would be necessary on renormalizability grounds. The technical 
issues of accommodating this within the LCO formalism may not be insurmountable
but would require new insights to find the effective potential.

In terms of future directions given that the core renormalization group
functions of QCD are known at five loops in the $\MSbar$ scheme, 
\cite{55,56,57,58}, the techniques are already available in principle to deduce
the next term in the series for $\zeta(a)$. This would require a five loop
computation of the $J$ correlation function. In addition the approach of
\cite{57} to renormalize QCD using a vacuum bubble expansion means that all the
core four loop Feynman integrals that are necessary to compute the four loop
graphs of the effective potential are known. Therefore extending our analysis
to the next loop order would seem viable in the foreseeable future. From 
another point of view it would be interesting to apply the LCO Lagrangian to
other problems such as the evaluation of $2$- and $3$-point functions akin to
that studied in \cite{37,40,41,42}. The reasoning would be to see if the mass
estimate extracted here was consistent with lattice data as well as testing
the effect the extra interactions present in (\ref{lagsig}) have. Finally we
recall that one of the early observations that led to the interest in studying
the value of $\langle \frac{1}{2} { A_\mu^a }^2 \rangle$ in the Landau gauge
arose in lattice and other studies, \cite{94,95,96,97,98}. It was noted that 
$O(1/p^2)$ power corrections arose in operator product expansion measurements. 
This ran counter to expectations that the leading correction would be 
$O(1/(p^2)^2)$. With the general acceptance that an effective gluon mass lurks 
in Yang-Mills theory the LCO Lagrangian might be the tool to formally 
re-examine the operator product expansion. 

\vspace{1cm}
\noindent
{\bf Acknowledgements.} This work was carried out with the support of STFC
through the Consolidated Grant ST/T000988/1. The diagrams were prepared with 
the {\sc Axodraw} package, \cite{99}. For the purpose of open access, the 
author has applied a Creative Commons Attribution (CC-BY) licence to any Author
Accepted Manuscript version arising. The data representing the main results
here are accessible in electronic form from the arXiv ancillary directory 
associated with the article.

\appendix

\sect{Expressions for arbitrary colour group.}

As the main analysis in the body of the article centred on the $SU(3)$ colour 
group and $\Nf$~$=$~$3$ we devote this section to recording the corresponding
expressions for an arbitrary group. First the counterterm associated with the 
source field is given to three loops by
\begin{eqnarray}
\delta \zeta &=&
\left[ 
-~ \frac{3}{2\epsilon}
+ \left[ 
[ 35 C_A - 16 \Nf T_F ] \frac{1}{8\epsilon^2} 
+ [ 32 \Nf T_F - 139 C_A ] \frac{1}{12\epsilon} \right] a 
\right. \nonumber \\
&& \left. ~
+ \left[
[ 584 C_A \Nf T_F - 665 C_A^2 - 128 \Nf^2 T_F^2 ] \frac{1}{48\epsilon^3} 
\right. \right. \nonumber \\
&& \left. \left. ~~~~~
+ [ 6629 C_A^2 - 4280 C_A \Nf T_F - 576 C_F \Nf T_F + 512 \Nf^2 T_F^2 ]
\frac{1}{144\epsilon^2} 
\right. \right. \nonumber \\
&& \left. \left. ~~~~~
+ [ 27648  \zeta_3C_A \Nf T_F 
+ 17524 C_A \Nf T_F 
- 6237 \zeta_3 C_A^2 
- 71551 C_A^2 
- 27648 \zeta_3 C_F \Nf T_F 
\right. \right. \nonumber \\
&& \left. \left. ~~~~~~~~~
+ 33120 C_F \Nf T_F 
+ 1280 \Nf^2 T_F^2 ] \frac{1}{864\epsilon} \right] a^2 
\right. \nonumber \\
&& \left. ~
+ \left[
[ 52535 C_A^3 
- 67416 C_A^2 \Nf T_F 
+ 28800 C_A \Nf^2 T_F^2
- 4096 \Nf^3 T_F^3 ] \frac{1}{1152\epsilon^4} 
\right. \right. \nonumber \\
&& \left. \left. ~~~~~
+ [ 1342920 C_A^2 \Nf T_F 
- 1274497 C_A^3 
+ 181728 C_A C_F \Nf T_F 
- 418176 C_A \Nf^2 T_F^2 
\right. \right. \nonumber \\
&& \left. \left. ~~~~~~~~~
- 78336 C_F \Nf^2 T_F^2
+ 32768 \Nf^3 T_F^3 ] \frac{1}{6912\epsilon^3}
\right. \right. \nonumber \\
&& \left. \left. ~~~~~
+ [ 488349 \zeta_3 C_A^3 
+ 8136934 C_A^3 
- 2243808 \zeta_3 C_A^2 \Nf T_F
- 5991276 C_A^2 \Nf T_F 
\right. \right. \nonumber \\
&& \left. \left. ~~~~~~~~~
+ 1997568 \zeta_3 C_A C_F \Nf T_F 
- 3516336 C_A C_F \Nf T_F
+ 884736 \zeta_3 C_A \Nf^2 T_F^2 
\right. \right. \nonumber \\
&& \left. \left. ~~~~~~~~~
+ 824256 C_A \Nf^2 T_F^2 
+ 31104 C_F^2 \Nf T_F
- 884736 \zeta_3 C_F \Nf^2 T_F^2 
+ 1301760 C_F \Nf^2 T_F^2
\right. \right. \nonumber \\
&& \left. \left. ~~~~~~~~~
+ 40960 \Nf^3 T_F^3 ] \frac{1}{20736\epsilon^2}
\right. \right. \nonumber \\
&& \left. \left. ~~~~~
+ [ 988362 \zeta_4 C_A^3 
- 7330584 \zeta_3 C_A^3 
+ 5416200 \zeta_5 C_A^3 
- 19251711 C_A^3
\right. \right. \nonumber \\
&& \left. \left. ~~~~~~~~~
+ 21050496 \zeta_3 C_A^2 \Nf T_F 
- 4860864 \zeta_4 C_A^2 \Nf T_F
- 4976640 \zeta_5 C_A^2 \Nf T_F 
\right. \right. \nonumber \\
&& \left. \left. ~~~~~~~~~
+ 7469896 C_A^2 \Nf T_F 
- 8861184 \zeta_3 C_A C_F \Nf T_F
+ 4492800 \zeta_4 C_A C_F \Nf T_F 
\right. \right. \nonumber \\
&& \left. \left. ~~~~~~~~~
- 11612160 \zeta_5 C_A C_F \Nf T_F
+ 18097952 C_A C_F \Nf T_F 
- 3723264 \zeta_3 C_A \Nf^2 T_F^2
\right. \right. \nonumber \\
&& \left. \left. ~~~~~~~~~
+ 1769472 \zeta_4 C_A \Nf^2 T_F^2 
- 278656 C_A \Nf^2 T_F^2
- 8460288 \zeta_3 C_F^2 \Nf T_F 
\right. \right. \nonumber \\
&& \left. \left. ~~~~~~~~~
+ 14376960 \zeta_5 C_F^2 \Nf T_F 
- 4405248 C_F^2 \Nf T_F
+ 3317760 \zeta_3 C_F \Nf^2 T_F^2 
\right. \right. \nonumber \\
&& \left. \left. ~~~~~~~~~
- 1769472 \zeta_4 C_F \Nf^2 T_F^2
- 1636864 C_F \Nf^2 T_F^2 
- 196608 \zeta_3 \Nf^3 T_F^3
\right. \right. \nonumber \\
&& \left. \left. ~~~~~~~~~
+ 65536 \Nf^3 T_F^3 ] \frac{1}{27648\epsilon} \right] a^3
\right] \NA ~+~ O(a^4) 
\end{eqnarray}
which is translated to its renormalization group function
\begin{eqnarray}
\delta(a) &=& \left[ -~ 3 
+ [ 32 \Nf T_F - 139 C_A ] \frac{a}{3}
\right. \nonumber \\
&& \left. ~
+ \left[ 
27648 \zeta_3 C_A \Nf T_F
- 6237 \zeta_3 C_A^2 
- 71551 C_A^2 
+ 17524 C_A \Nf T_F 
\right. \right. \nonumber \\
&& \left. \left. ~~~~~
- 27648 \zeta_3 C_F \Nf T_F 
+ 33120 C_F \Nf T_F
+ 1280 \Nf^2 T_F^2 \right] \frac{a^2}{144}  
\right. \nonumber \\
&& \left. ~
+ \left[ 
5416200 \zeta_5 C_A^3
+ 988362 \zeta_4 C_A^3 
- 7330584 \zeta_3 C_A^3 
- 19251711 C_A^3 
\right. \right. \nonumber \\
&& \left. \left. ~~~~~
+ 21050496 \zeta_3 C_A^2 \Nf T_F
- 4860864 \zeta_4 C_A^2 \Nf T_F 
- 4976640 \zeta_5 C_A^2 \Nf T_F
\right. \right. \nonumber \\
&& \left. \left. ~~~~~
+ 7469896 C_A^2 \Nf T_F 
- 8861184 \zeta_3 C_A C_F \Nf T_F
+ 4492800 \zeta_4 C_A C_F \Nf T_F 
\right. \right. \nonumber \\
&& \left. \left. ~~~~~
- 11612160 \zeta_5 C_A C_F \Nf T_F
+ 18097952 C_A C_F \Nf T_F 
- 3723264 \zeta_3 C_A \Nf^2 T_F^2
\right. \right. \nonumber \\
&& \left. \left. ~~~~~
+ 1769472 \zeta_4 C_A \Nf^2 T_F^2 
- 278656 C_A \Nf^2 T_F^2
- 8460288 \zeta_3 C_F^2 \Nf T_F 
\right. \right. \nonumber \\
&& \left. \left. ~~~~~
+ 14376960 \zeta_5 C_F^2 \Nf T_F
- 4405248 C_F^2 \Nf T_F 
+ 3317760 \zeta_3 C_F \Nf^2 T_F^2
\right. \right. \nonumber \\
&& \left. \left. ~~~~~
- 1769472 \zeta_4 C_F \Nf^2 T_F^2 
- 1636864 C_F \Nf^2 T_F^2
- 196608 \zeta_3 \Nf^3 T_F^3 
\right. \right. \nonumber \\
&& \left. \left. ~~~~~
+ 65536 \Nf^3 T_F^3 \right] \frac{a^3}{3456}
\right] \NA ~+~ O(a^4) ~.
\end{eqnarray}
One reason for noting the counterterm was to give assurance that the derivation
of $\delta(a)$ was consistent with the renormalization group formalism. By this
we mean that the double and triple poles in $\epsilon$ are not independent but
determined by the simple poles at lower loop orders. Equipped with this the
expression for $\zeta(a)$ can be deduced from (\ref{zetasoln}) leading to
\begin{eqnarray}
\frac{1}{\zeta(a)} &=& 
\left[ 
\left[ 
\frac{13}{9} C_A
- \frac{8}{9} T_F \Nf
\right] a
\right. \nonumber \\
&& \left. ~
+ \left[ 
\frac{105}{4} C_F C_A^2 \lambda_2
- \frac{8129}{1296} C_A^2
- \frac{105}{16} C_A^3 \lambda_2
- \frac{3}{4} C_F C_A
+ \frac{557}{81} T_F C_A \Nf
- \frac{4}{3} T_F C_F \Nf
\right. \right. \nonumber \\
&& \left. \left. ~~~~~~
- \frac{128}{81} T_F^2 \Nf^2
\right] a^2
\right. \nonumber \\
&& \left. ~
+ \left[ 
\frac{1459}{36} C_A^4 \lambda_3
- \frac{15005}{1728} C_A^3
- \frac{4795}{128} C_A^4 \lambda_2
- \frac{11025}{128} C_A^5 \lambda_2^2
- \frac{629}{144} C_F C_A^2
\right. \right. \nonumber \\
&& \left. \left. ~~~~~~
+ \frac{950}{9} C_F C_A^3 \lambda_3
- \frac{245}{4} C_F C_A^3 \lambda_2
+ \frac{11025}{16} C_F C_A^4 \lambda_2^2
+ \frac{8}{3} C_F^2 C_A
\right. \right. \nonumber \\
&& \left. \left. ~~~~~~
- \frac{1463}{3} C_F^2 C_A^2 \lambda_3
+ \frac{6755}{8} C_F^2 C_A^2 \lambda_2
- \frac{11025}{8} C_F^2 C_A^3 \lambda_2^2
- \frac{20275}{1944} T_F C_A^2 \Nf
\right. \right. \nonumber \\
&& \left. \left. ~~~~~~
+ \frac{2311}{81} T_F C_F C_A \Nf
+ \frac{10}{9} T_F C_F^2 \Nf
+ \frac{3674}{243} T_F^2 C_A \Nf^2
- \frac{1880}{81} T_F^2 C_F \Nf^2
\right. \right. \nonumber \\
&& \left. \left. ~~~~~~
- \frac{896}{243} T_F^3 \Nf^3
- \frac{1661}{144} \zeta_3 C_A^3
+ \frac{763}{6} \zeta_3 C_A^4 \lambda_3
+ \frac{20}{3} \zeta_3 C_F C_A^2
- \frac{380}{3} \zeta_3 C_F C_A^3 \lambda_3
\right. \right. \nonumber \\
&& \left. \left. ~~~~~~
+ \frac{1343}{54} \zeta_3 T_F C_A^2 \Nf
- \frac{592}{27} \zeta_3 T_F C_F C_A \Nf
- \frac{512}{27} \zeta_3 T_F^2 C_A \Nf^2
+ \frac{512}{27} \zeta_3 T_F^2 C_F \Nf^2
\right] a^3
\right. \nonumber \\
&& \left. ~
+ \left[ 
\frac{256}{27} \Nf \frac{d_F^{abcd} d_A^{abcd}}{\NA}
- \frac{659}{864} \frac{d_A^{abcd} d_A^{abcd}}{\NA}
- \frac{352}{27} \Nf^2 \frac{d_F^{abcd} d_F^{abcd}}{\NA}
+ \frac{659}{32} C_A \lambda_4 \frac{d_A^{abcd} d_A^{abcd}}{\NA}
\right. \right. \nonumber \\
&& \left. \left. ~~~~~~
- 256 C_A \Nf \lambda_4 \frac{d_F^{abcd} d_A^{abcd}}{\NA}
+ 352 C_A \Nf^2 \lambda_4 \frac{d_F^{abcd} d_F^{abcd}}{\NA}
- \frac{5778875537}{53747712} C_A^4
\right. \right. \nonumber \\
&& \left. \left. ~~~~~~
- \frac{19964625}{8192} C_A^5 \lambda_4
+ \frac{24803}{36} C_A^5 \lambda_3
+ \frac{176715}{1024} C_A^5 \lambda_2
- \frac{305025}{256} C_A^6 \lambda_2^2
\right. \right. \nonumber \\
&& \left. \left. ~~~~~~
- \frac{1157625}{1024} C_A^7 \lambda_2^3
+ \frac{14933}{1024} C_F C_A^3
- \frac{20457603}{1024} C_F C_A^4 \lambda_4
+ \frac{67735}{12} C_F C_A^4 \lambda_3
\right. \right. \nonumber \\
&& \left. \left. ~~~~~~
- \frac{1689905}{768} C_F C_A^4 \lambda_2
+ \frac{591675}{128} C_F C_A^5 \lambda_2^2
+ \frac{3472875}{256} C_F C_A^6 \lambda_2^3
+ \frac{48155}{2304} C_F^2 C_A^2
\right. \right. \nonumber \\
&& \left. \left. ~~~~~~
+ \frac{550551}{256} C_F^2 C_A^3 \lambda_4
+ \frac{15637}{9} C_F^2 C_A^3 \lambda_3
- \frac{520135}{96} C_F^2 C_A^3 \lambda_2
+ \frac{40425}{2} C_F^2 C_A^4 \lambda_2^2
\right. \right. \nonumber \\
&& \left. \left. ~~~~~~
- \frac{3472875}{64} C_F^2 C_A^5 \lambda_2^3
- \frac{203}{96} C_F^3 C_A
+ \frac{3016773}{32} C_F^3 C_A^2 \lambda_4
- \frac{138985}{3} C_F^3 C_A^2 \lambda_3
\right. \right. \nonumber \\
&& \left. \left. ~~~~~~
+ \frac{1100015}{24} C_F^3 C_A^2 \lambda_2
- \frac{628425}{8} C_F^3 C_A^3 \lambda_2^2
+ \frac{1157625}{16} C_F^3 C_A^4 \lambda_2^3
\right. \right. \nonumber \\
&& \left. \left. ~~~~~~
+ \frac{215810801}{1679616} T_F C_A^3 \Nf
+ \frac{3323423}{23328} T_F C_F C_A^2 \Nf
- \frac{75553}{648} T_F C_F^2 C_A \Nf
+ 6 T_F C_F^3 \Nf
\right. \right. \nonumber \\
&& \left. \left. ~~~~~~
- \frac{1759765}{17496} T_F^2 C_A^2 \Nf^2
- \frac{22360}{729} T_F^2 C_F C_A \Nf^2
+ \frac{6940}{81} T_F^2 C_F^2 \Nf^2
+ \frac{310736}{6561} T_F^3 C_A \Nf^3
\right. \right. \nonumber \\
&& \left. \left. ~~~~~~
- \frac{33488}{729} T_F^3 C_F \Nf^3
- \frac{62464}{6561} T_F^4 \Nf^4
+ \frac{3395}{128} \zeta_5 \frac{d_A^{abcd} d_A^{abcd}}{\NA}
- 20 \zeta_5 \Nf \frac{d_F^{abcd} d_A^{abcd}}{\NA}
\right. \right. \nonumber \\
&& \left. \left. ~~~~~~
- \frac{91665}{128} \zeta_5 C_A \lambda_4 \frac{d_A^{abcd} d_A^{abcd}}{\NA}
+ 540 \zeta_5 C_A \Nf \lambda_4 \frac{d_F^{abcd} d_A^{abcd}}{\NA}
+ \frac{4608325}{27648} \zeta_5 C_A^4
\right. \right. \nonumber \\
&& \left. \left. ~~~~~~
- \frac{2612565}{1024} \zeta_5 C_A^5 \lambda_4
+ \frac{2565}{16} \zeta_5 C_F C_A^3
- \frac{202635}{16} \zeta_5 C_F C_A^4 \lambda_4
- \frac{1485}{8} \zeta_5 C_F^2 C_A^2
\right. \right. \nonumber \\
&& \left. \left. ~~~~~~
+ \frac{117315}{8} \zeta_5 C_F^2 C_A^3 \lambda_4
- \frac{47035}{216} \zeta_5 T_F C_A^3 \Nf
- \frac{1930}{9} \zeta_5 T_F C_F C_A^2 \Nf
\right. \right. \nonumber \\
&& \left. \left. ~~~~~~
+ \frac{7580}{27} \zeta_5 T_F C_F^2 C_A \Nf
+ \frac{320}{3} \zeta_5 T_F^2 C_A^2 \Nf^2
+ \frac{2240}{9} \zeta_5 T_F^2 C_F C_A \Nf^2
\right. \right. \nonumber \\
&& \left. \left. ~~~~~~
- \frac{8320}{27} \zeta_5 T_F^2 C_F^2 \Nf^2
+ \frac{13247}{576} \zeta_4 C_A^4
- \frac{7625}{54} \zeta_4 T_F C_A^3 \Nf
+ \frac{3172}{27} \zeta_4 T_F C_F C_A^2 \Nf
\right. \right. \nonumber \\
&& \left. \left. ~~~~~~
+ \frac{3775}{27} \zeta_4 T_F^2 C_A^2 \Nf^2
- \frac{3616}{27} \zeta_4 T_F^2 C_F C_A \Nf^2
- \frac{1024}{27} \zeta_4 T_F^3 C_A \Nf^3
+ \frac{1024}{27} \zeta_4 T_F^3 C_F \Nf^3
\right. \right. \nonumber \\
&& \left. \left. ~~~~~~
+ \frac{989}{72} \zeta_3 \frac{d_A^{abcd} d_A^{abcd}}{\NA}
- \frac{688}{9} \zeta_3 \Nf \frac{d_F^{abcd} d_A^{abcd}}{\NA}
+ \frac{256}{9} \zeta_3 \Nf^2 \frac{d_F^{abcd} d_F^{abcd}}{\NA}
\right. \right. \nonumber \\
&& \left. \left. ~~~~~~
- \frac{2967}{8} \zeta_3 C_A \lambda_4 \frac{d_A^{abcd} d_A^{abcd}}{\NA}
+ 2064 \zeta_3 C_A \Nf \lambda_4 \frac{d_F^{abcd} d_A^{abcd}}{\NA}
\right. \right. \nonumber \\
&& \left. \left. ~~~~~~
- 768 \zeta_3 C_A \Nf^2 \lambda_4 \frac{d_F^{abcd} d_F^{abcd}}{\NA}
- \frac{21102833}{248832} \zeta_3 C_A^4
- \frac{14781555}{1024} \zeta_3 C_A^5 \lambda_4
\right. \right. \nonumber \\
&& \left. \left. ~~~~~~
+ \frac{12971}{6} \zeta_3 C_A^5 \lambda_3
+ \frac{36435}{32} \zeta_3 C_A^5 \lambda_2
- \frac{16405}{128} \zeta_3 C_F C_A^3
- \frac{2339901}{128} \zeta_3 C_F C_A^4 \lambda_4
\right. \right. \nonumber \\
&& \left. \left. ~~~~~~
+ \frac{19855}{2} \zeta_3 C_F C_A^4 \lambda_3
- \frac{91245}{16} \zeta_3 C_F C_A^4 \lambda_2
+ \frac{4795}{48} \zeta_3 C_F^2 C_A^2
+ \frac{508365}{16} \zeta_3 C_F^2 C_A^3 \lambda_4
\right. \right. \nonumber \\
&& \left. \left. ~~~~~~
- \frac{36100}{3} \zeta_3 C_F^2 C_A^3 \lambda_3
+ \frac{18375}{4} \zeta_3 C_F^2 C_A^3 \lambda_2
+ \frac{890945}{2592} \zeta_3 T_F C_A^3 \Nf
\right. \right. \nonumber \\
&& \left. \left. ~~~~~~
- \frac{13775}{972} \zeta_3 T_F C_F C_A^2 \Nf
- \frac{4378}{27} \zeta_3 T_F C_F^2 C_A \Nf
- \frac{45020}{243} \zeta_3 T_F^2 C_A^2 \Nf^2
\right. \right. \nonumber \\
&& \left. \left. ~~~~~~
- \frac{16276}{243} \zeta_3 T_F^2 C_F C_A \Nf^2
+ \frac{4928}{27} \zeta_3 T_F^2 C_F^2 \Nf^2
+ \frac{640}{81} \zeta_3 T_F^3 C_A \Nf^3
- \frac{896}{243} \zeta_3 T_F^3 C_F \Nf^3
\right. \right. \nonumber \\
&& \left. \left. ~~~~~~
+ \frac{1024}{243} \zeta_3 T_F^4 \Nf^4
\right] a^4
\right] \frac{1}{\NA} ~+~ O(a^5) ~.
\label{zetagen}
\end{eqnarray}
This is clearly a more involved expression compared to that of the previous
order given in \cite{51,52}. The reason for this is mainly due to the four loop
terms of the $\beta$-function and the gluon mass anomalous dimensions, 
\cite{58,100} which both involve rank $4$ colour Casimirs which are not present 
at lower loop order. These are based on the totally symmetric tensor,
\cite{74},
\begin{equation}
d_R^{abcd} ~=~ \frac{1}{6} \mbox{Tr} \left( T_R^a T_R^{(b} T_R^c T_R^{d)}
\right)
\end{equation}
for an arbitrary group representation. The designation of $R$ to be $F$ or $A$
indicates evaluation in either the fundamental or adjoint representations. Next
we have 
\begin{eqnarray}
Z_\zeta^{-1} &=& 1
+ \left[
\frac{13}{6} C_A
- \frac{4}{3} T_F \Nf
\right] 
\frac{a}{\epsilon}
\nonumber \\
&& 
+ \left[ 
\left[
T_F C_A \Nf
- \frac{13}{8} C_A^2
\right] 
\frac{1}{\epsilon^2}  
\right. \nonumber \\
&& \left. ~~~~
+ \left[
\frac{703}{96} C_A^2
- \frac{315}{32} C_A^3 \lambda_2
- \frac{9}{8} C_F C_A
+ \frac{315}{8} C_F C_A^2 \lambda_2
- \frac{23}{6} T_F C_A \Nf
\right. \right. \nonumber \\
&& \left. \left. ~~~~~~~~
- 2 T_F C_F \Nf
\right] 
\frac{1}{\epsilon}  
\right] a^2
\nonumber \\
&& 
+ \left[
\left[
\frac{403}{144} C_A^3
- \frac{22}{9} T_F C_A^2 \Nf
+ \frac{4}{9} T_F^2 C_A \Nf^2
\right] 
\frac{1}{\epsilon^3}
\right. \nonumber \\
&& \left. ~~~~
+ \left[
\frac{945}{64} C_A^4 \lambda_2
- \frac{521}{64} C_A^3
+ \frac{27}{16} C_F C_A^2
- \frac{945}{16} C_F C_A^3 \lambda_2
+ \frac{61}{12} T_F C_A^2 \Nf
\right. \right. \nonumber \\
&& \left. \left. ~~~~~~~~
+ \frac{40}{9} T_F C_F C_A \Nf
- \frac{2}{3} T_F^2 C_A \Nf^2
- \frac{8}{9} T_F^2 C_F \Nf^2
\right] 
\frac{1}{\epsilon^2}
\right. \nonumber \\
&& \left. ~~~~
+ \left[
\frac{113515}{3456} C_A^3
+ \frac{1459}{24} C_A^4 \lambda_3
- \frac{24045}{256} C_A^4 \lambda_2
- \frac{33075}{256} C_A^5 \lambda_2^2
- \frac{1043}{96} C_F C_A^2
\right. \right. \nonumber \\
&& \left. \left. ~~~~~~~~
+ \frac{475}{3} C_F C_A^3 \lambda_3
+ \frac{945}{16} C_F C_A^3 \lambda_2
+ \frac{33075}{32} C_F C_A^4 \lambda_2^2
+ 4 C_F^2 C_A
- \frac{1463}{2} C_F^2 C_A^2 \lambda_3
\right. \right. \nonumber \\
&& \left. \left. ~~~~~~~~
+ \frac{20265}{16} C_F^2 C_A^2 \lambda_2
- \frac{33075}{16} C_F^2 C_A^3 \lambda_2^2
- \frac{3193}{144} T_F C_A^2 \Nf
- \frac{1405}{54} T_F C_F C_A \Nf
\right. \right. \nonumber \\
&& \left. \left. ~~~~~~~~
+ \frac{5}{3} T_F C_F^2 \Nf
+ \frac{52}{27} T_F^2 C_A \Nf^2
+ \frac{76}{27} T_F^2 C_F \Nf^2
- \frac{55}{8} \zeta_3 C_A^3
+ \frac{763}{4} \zeta_3 C_A^4 \lambda_3
\right. \right. \nonumber \\
&& \left. \left. ~~~~~~~~
+ 10 \zeta_3 C_F C_A^2
- 190 \zeta_3 C_F C_A^3 \lambda_3
- \frac{46}{3} \zeta_3 T_F C_A^2 \Nf
+ \frac{40}{3} \zeta_3 T_F C_F C_A \Nf
\right] 
\frac{1}{\epsilon}
\right] a^3 \nonumber \\
&& +~ O(a^4)
\end{eqnarray}
which was defined in (\ref{Zzdef}). Another quantity required prior to 
determining the effective potential is the renormalization constant 
\begin{eqnarray}
Z_m &=& 1 + \left[ \frac{4}{3} T_F \Nf - \frac{35}{12} C_A \right]
\frac{a}{\epsilon} \nonumber \\
&& 
+ \left[ \left[ \frac{2765}{288} C_A^2 + \frac{16}{9} T_F^2 \Nf^2
- \frac{149}{18} T_F \Nf C_A \right] \frac{1}{\epsilon^2}
\right. \nonumber \\
&& \left. ~~~
+ \left[ 2 T_F \Nf C_F - \frac{449}{96} C_A^2
+ \frac{35}{12} T_F \Nf C_A \right] \frac{1}{\epsilon}
\right] a^2 ~+~ O(a^3) ~.
\end{eqnarray}
While this is already available to high loop order via the Slavnov-Taylor
identity, \cite{59,61}, it is provided in the conventions used to derive 
$V(\sigma)$ and is the source for deducing $\gamma_m(a)$.

Finally the full effective potential at three loops for the Landau gauge gluon 
mass operator for an arbitrary colour group is 
\begin{eqnarray}
V(\sigma) &=&
\left[
\frac{9}{2} \lambda_1
+ \left[
\frac{105}{2} C_F \lambda_2
- \frac{39}{2} C_F \lambda_1
- \frac{13}{8}
- \frac{105}{8} C_A \lambda_2
- \frac{9}{4} C_A \lambda_1
+ \frac{3}{4} \overline{\ln}(\bar{g} \bar{\sigma})
\right] a 
\right. \nonumber \\
&& \left.
+ \left[
\frac{1459}{32} C_A^2 \lambda_3
- \frac{7665}{64} C_A^2 \lambda_2
- \frac{2373}{64} C_A^2 \lambda_1
- \frac{593}{64} C_A
- \frac{11025}{64} C_A^3 \lambda_2^2
- \frac{247}{16} C_F
\right. \right. \nonumber \\
&& \left. \left. ~~~~
+ \frac{475}{4} C_F C_A \lambda_3
+ 210 C_F C_A \lambda_2
+ \frac{429}{16} C_F C_A \lambda_1
+ \frac{11025}{8} C_F C_A^2 \lambda_2^2
- \frac{4389}{8} C_F^2 \lambda_3
\right. \right. \nonumber \\
&& \left. \left. ~~~~
+ \frac{4305}{4} C_F^2 \lambda_2
+ \frac{39}{8} C_F^2 \lambda_1
- \frac{11025}{4} C_F^2 C_A \lambda_2^2
- 12 \zeta_3 C_A
+ \frac{2289}{16} \zeta_3 C_A^2 \lambda_3
\right. \right. \nonumber \\
&& \left. \left. ~~~~
+ \frac{1377}{32} \zeta_3 C_A^2 \lambda_1
+ 12 \zeta_3 C_F
- \frac{285}{2} \zeta_3 C_F C_A \lambda_3
- \frac{117}{2} \zeta_3 C_F C_A \lambda_1
- \frac{1}{16} \zeta_2 C_A
\right. \right. \nonumber \\
&& \left. \left. ~~~~
+ 2 \zeta_2 T_F \Nf
+ \frac{99}{16} \sqrt{3} \mbox{Ls}_2(\twopithree) C_A
\right. \right. \nonumber \\
&& \left. \left. ~~~~
+ \left[ 
\frac{75}{16} C_A
+ \frac{315}{16} C_A^2 \lambda_2
+ \frac{9}{4} C_F
- \frac{315}{4} C_F C_A \lambda_2
\right]
\overline{\ln}(\bar{g} \bar{\sigma}) 
- \frac{9}{16} C_A \overline{\ln}^2(\bar{g} \bar{\sigma})
\right] a^2 
\right. \nonumber \\
&& \left.
+ \left[
\left[
\frac{274995}{64} \zeta_5 
+ \frac{8901}{4} \zeta_3 
- \frac{1977}{16} 
\right] \lambda_1 \lambda_4 \frac{d_A^{abcd} d_A^{abcd}}{\NA}
\right. \right. \nonumber \\
&& \left. \left. ~~~~
+ \left[
1536
- 3240 
- 12384 \zeta_3
\right] \Nf \lambda_1 \lambda_4 \frac{d_F^{abcd} d_A^{abcd}}{\NA}
\right. \right. \nonumber \\
&& \left. \left. ~~~~
+ \left[
4608 \zeta_3 
- 2112 
\right] \Nf^2 \lambda_1 \lambda_4 \frac{d_F^{abcd} d_F^{abcd}}{\NA}
\right. \right. \nonumber \\
&& \left. \left. ~~~~
- \frac{4647721}{27648} C_A^2
- \frac{6654875}{3072} C_A^3 \lambda_4
+ \frac{77327}{96} C_A^3 \lambda_3
- \frac{162085}{384} C_A^3 \lambda_2
- \frac{64193}{384} C_A^3 \lambda_1
\right. \right. \nonumber \\
&& \left. \left. ~~~~
- \frac{738675}{256} C_A^4 \lambda_2^2
- \frac{1157625}{512} C_A^5 \lambda_2^3
- \frac{6301}{48} C_F C_A
- \frac{2273067}{128} C_F C_A^2 \lambda_4
\right. \right. \nonumber \\
&& \left. \left. ~~~~
+ \frac{617215}{96} C_F C_A^2 \lambda_3
- \frac{56245}{64} C_F C_A^2 \lambda_2
+ \frac{74087}{192} C_F C_A^2 \lambda_1
+ \frac{488775}{32} C_F C_A^3 \lambda_2^2
\right. \right. \nonumber \\
&& \left. \left. ~~~~
+ \frac{3472875}{128} C_F C_A^4 \lambda_2^3
+ \frac{1791}{32} C_F^2
+ \frac{183517}{96} C_F^2 C_A \lambda_4
+ \frac{12711}{8} C_F^2 C_A \lambda_3
\right. \right. \nonumber \\
&& \left. \left. ~~~~
- \frac{168315}{32} C_F^2 C_A \lambda_2
- \frac{7579}{24} C_F^2 C_A \lambda_1
+ \frac{260925}{16} C_F^2 C_A^2 \lambda_2^2
- \frac{3472875}{32} C_F^2 C_A^3 \lambda_2^3
\right. \right. \nonumber \\
&& \left. \left. ~~~~
+ \frac{335197}{4} C_F^3 \lambda_4
- \frac{416955}{8} C_F^3 \lambda_3
+ \frac{1490755}{24} C_F^3 \lambda_2
+ \frac{1963}{24} C_F^3 \lambda_1
\right. \right. \nonumber \\
&& \left. \left. ~~~~
- 124950 C_F^3 C_A \lambda_2^2
+ \frac{1157625}{8} C_F^3 C_A^2 \lambda_2^3
+ \frac{29257}{864} T_F C_A \Nf
+ \frac{1903}{24} T_F C_F \Nf
\right. \right. \nonumber \\
&& \left. \left. ~~~~
+ \frac{135}{2} \zeta_5 C_A^2
- \frac{290285}{128} \zeta_5 C_A^3 \lambda_4
- \frac{43855}{512} \zeta_5 C_A^3 \lambda_1
+ \frac{315}{2} \zeta_5 C_F C_A
\right. \right. \nonumber \\
&& \left. \left. ~~~~
- \frac{22515}{2} \zeta_5 C_F C_A^2 \lambda_4
- 195 \zeta_5 C_F C_A^2 \lambda_1
- 195 \zeta_5 C_F^2
+ 13035 \zeta_5 C_F^2 C_A \lambda_4
\right. \right. \nonumber \\
&& \left. \left. ~~~~
+ 390 \zeta_5 C_F^2 C_A \lambda_1
+ \frac{20913}{512} \zeta_4 C_A^2
- \frac{4131}{64} \zeta_4 C_A^3 \lambda_1
- \frac{27}{4} \zeta_4 C_F C_A
+ \frac{351}{4} \zeta_4 C_F C_A^2 \lambda_1
\right. \right. \nonumber \\
&& \left. \left. ~~~~
- 88 \zeta_4 T_F C_A \Nf
+ 108 \zeta_4 T_F C_F \Nf
- \frac{12703}{192} \zeta_3 C_A^2
- \frac{1642395}{128} \zeta_3 C_A^3 \lambda_4
\right. \right. \nonumber \\
&& \left. \left. ~~~~
+ \frac{40439}{16} \zeta_3 C_A^3 \lambda_3
+ \frac{36435}{16} \zeta_3 C_A^3 \lambda_2
+ \frac{67125}{256} \zeta_3 C_A^3 \lambda_1
- \frac{1439}{32} \zeta_3 C_F C_A
\right. \right. \nonumber \\
&& \left. \left. ~~~~
- \frac{259989}{16} \zeta_3 C_F C_A^2 \lambda_4
+ \frac{177175}{16} \zeta_3 C_F C_A^2 \lambda_3
- \frac{91245}{8} \zeta_3 C_F C_A^2 \lambda_2
- \frac{2613}{32} \zeta_3 C_F C_A^2 \lambda_1
\right. \right. \nonumber \\
&& \left. \left. ~~~~
+ \frac{231}{2} \zeta_3 C_F^2
+ \frac{56485}{2} \zeta_3 C_F^2 C_A \lambda_4
- \frac{27075}{2} \zeta_3 C_F^2 C_A \lambda_3
+ \frac{18375}{2} \zeta_3 C_F^2 C_A \lambda_2
\right. \right. \nonumber \\
&& \left. \left. ~~~~
- 299 \zeta_3 C_F^2 C_A \lambda_1
+ \frac{241}{6} \zeta_3 T_F C_A \Nf
- \frac{263}{3} \zeta_3 T_F C_F \Nf
- \frac{9883}{2304} \zeta_2 C_A^2
+ \frac{7245}{64} \zeta_2 C_A^3 \lambda_2
\right. \right. \nonumber \\
&& \left. \left. ~~~~
+ \frac{207}{16} \zeta_2 C_F C_A
- \frac{7245}{16} \zeta_2 C_F C_A^2 \lambda_2
+ \frac{6701}{288} \zeta_2 T_F C_A \Nf
- 16 \zeta_2 T_F C_F \Nf
\right. \right. \nonumber \\
&& \left. \left. ~~~~
- \frac{16}{9} \zeta_2 T_F^2 \Nf^2
- \frac{225}{8} \mbox{Li}_4(\half) C_A^2
+ 64 \mbox{Li}_4(\half) T_F C_A \Nf
- 128 \mbox{Li}_4(\half) T_F C_F \Nf
\right. \right. \nonumber \\
&& \left. \left. ~~~~
+ \frac{4239}{256} \mbox{Ls}^2_2(\twopithree) C_A^2
- 36 \mbox{Ls}^2_2(\twopithree) T_F C_A \Nf
+ \frac{225}{32} \ln^2(2) \zeta_2 C_A^2
- 16 \ln^2(2) \zeta_2 T_F C_A \Nf
\right. \right. \nonumber \\
&& \left. \left. ~~~~
+ 32 \ln^2(2) \zeta_2 T_F C_F \Nf
- \frac{75}{64} \ln^4(2) C_A^2
+ \frac{8}{3} \ln^4(2) T_F C_A \Nf
- \frac{16}{3} \ln^4(2) T_F C_F \Nf
\right. \right. \nonumber \\
&& \left. \left. ~~~~
- \frac{11}{432} \sqrt{3} \pi^3 C_A^2
+ \frac{11}{18} \sqrt{3} \pi^3 T_F C_A \Nf
+ \frac{677}{64} \sqrt{3} \mbox{Ls}_2(\twopithree) C_A^2
+ 11 \sqrt{3} \mbox{Ls}_2(\twopithree) T_F C_A \Nf
\right. \right. \nonumber \\
&& \left. \left. ~~~~
+ \frac{10395}{64} \sqrt{3} \mbox{Ls}_2(\twopithree) C_A^3 \lambda_2
+ \frac{297}{16} \sqrt{3} \mbox{Ls}_2(\twopithree) C_F C_A
- \frac{10395}{16} \sqrt{3} \mbox{Ls}_2(\twopithree) C_F C_A^2 \lambda_2
\right. \right. \nonumber \\
&& \left. \left. ~~~~
+ \left[
 \frac{61787}{768} C_A^2
- \frac{1459}{16} C_A^3 \lambda_3
+ \frac{34965}{128} C_A^3 \lambda_2
+ \frac{99225}{256} C_A^4 \lambda_2^2
+ \frac{979}{32} C_F C_A
\right. \right. \right. \nonumber \\
&& \left. \left. \left. ~~~~~~~
- \frac{475}{2} C_F C_A^2 \lambda_3
- \frac{8715}{16} C_F C_A^2 \lambda_2
- \frac{99225}{32} C_F C_A^3 \lambda_2^2
- \frac{3}{16} C_F^2
\right. \right. \right. \nonumber \\
&& \left. \left. \left. ~~~~~~~
+ \frac{4389}{4} C_F^2 C_A \lambda_3
- \frac{17535}{8} C_F^2 C_A \lambda_2
+ \frac{99225}{16} C_F^2 C_A^2 \lambda_2^2
- \frac{307}{16} T_F C_A \Nf
\right. \right. \right. \nonumber \\
&& \left. \left. \left. ~~~~~~~
- \frac{70}{3} T_F C_F \Nf
+ \frac{1335}{64} \zeta_3 C_A^2
- \frac{2289}{8} \zeta_3 C_A^3 \lambda_3
- 15 \zeta_3 C_F C_A
+ 285 \zeta_3 C_F C_A^2 \lambda_3
\right. \right. \right. \nonumber \\
&& \left. \left. \left. ~~~~~~~
- 16 \zeta_3 T_F C_A \Nf
+ 16 \zeta_3 T_F C_F \Nf
- \frac{341}{384} \zeta_2 C_A^2
- \frac{485}{48} \zeta_2 T_F C_A \Nf
+ \frac{8}{3} \zeta_2 T_F^2 \Nf^2
\right. \right. \right. \nonumber \\
&& \left. \left. \left. ~~~~~~~
- \frac{1023}{32} \sqrt{3} \mbox{Ls}_2(\twopithree) C_A^2
+ \frac{33}{4} \sqrt{3} \mbox{Ls}_2(\twopithree) T_F C_A \Nf
\right]
\overline{\ln}(\bar{g} \bar{\sigma})
\right. \right. \nonumber \\
&& \left. \left. ~~~~
+ \left[ 
\frac{2835}{32} C_F C_A^2 \lambda_2
+ \frac{15}{4} T_F C_A \Nf
+ \frac{3}{2} T_F C_F \Nf
- \frac{1791}{128} C_A^2
- \frac{2835}{128} C_A^3 \lambda_2
\right. \right. \right. \nonumber \\
&& \left. \left. \left. ~~~~~~~
- \frac{81}{32} C_F C_A
\right] \overline{\ln}^2(\bar{g} \bar{\sigma}) 
+ \left[ 
\frac{31}{32} C_A^2
- \frac{1}{4} T_F C_A \Nf
\right]
\overline{\ln}^3(\bar{g} \bar{\sigma})
\right] a^3 
\right] \NA \bar{\sigma}^2 \nonumber \\
&& +~ O(a^4) ~.
\label{effpotarb}
\end{eqnarray}
We have used the same shorthand notation as \cite{52} for various colour group 
combinations but appended these with a new one, $\lambda_4$, that arises at 
this new loop order. These are
\begin{eqnarray}
\lambda_1 &=& \frac{1}{[13 C_A-8 T_F \Nf]} ~~,~~ 
\lambda_2 ~=~ \frac{1}{[35 C_A-16 T_F \Nf]} \nonumber \\ 
\lambda_3 &=& \frac{1}{[19 C_A-8 T_F \Nf]} ~~,~~ 
\lambda_4 ~=~ \frac{1}{[79 C_A-32 T_F \Nf]} ~.
\end{eqnarray}
For completeness we note, \cite{74},
\begin{eqnarray}
d_F^{abcd} d_F^{abcd} &=& \frac{(\Nc^2-1)(\Nc^4-6\Nc^2+18)}{96\Nc^2} ~~~,~~~ 
d_F^{abcd} d_A^{abcd} ~=~ \frac{\Nc(\Nc^2-1)(\Nc^2+6)}{48} \nonumber \\
d_A^{abcd} d_A^{abcd} &=& \frac{\Nc^2(\Nc^2-1)(\Nc^2+36)}{24}
\end{eqnarray}
for $SU(\Nc)$.

\end{document}